\documentclass[a4paper,12pt]{article}
\usepackage{amsmath}
\usepackage{amssymb}
\usepackage{graphicx}
\usepackage[usenames]{color}
\usepackage{latexsym}
\usepackage{mathrsfs}
\definecolor{MyRed}{cmyk}{0,1,1,0.15}

\def\parn              {  \par\noindent }

\def\al{\alpha}
\def\be{\beta}
\def\ga{\gamma} 
\def\ep{\epsilon}
\def\lam{\lambda}

\def\calG{{\cal G}} \def\calH{{\cal H}} 
\def\calJ{{\cal J}} \def\calK{{\cal K}} \def\calL{{\cal L}}
  \def\calO{{\cal O}}
  \def\calR{{\cal R}}

\def\vecii#1#2      {  \left(\begin{array}{c}#1\\#2\end{array}\right)  }
\def\veciii#1#2#3   {  \left(\begin{array}{c}#1\\#2\\#3\end{array}
                     \right)  }
\def\veciv#1#2#3#4  {  \left(\begin{array}{c}#1\\#2\\#3\\#4
                                 \end{array}\right)  }
\def\vecfv#1#2#3#4#5 {  \left(\begin{array}{c}#1\\#2\\#3\\#4\\#5
                                 \end{array}\right)  }

\def\matrixii#1#2#3#4            {  \left(\begin{array}{cc}#1&#2\\#3&#4
                                       \end{array}\right) }
\def\matrixiii#1#2#3#4#5#6#7#8#9 {  \left(\begin{array}{ccc}#1&#2&#3\\
                                     #4&#5&#6\\#7&#8&#9\end{array}
                               \right)  }
\def\mativ#1#2#3#4               {  \left(\begin{array}{cccc}
                                       #1\\#2\\#3\\#4\end{array}\right) }
\def\matv#1#2#3#4#5              {  \left(\begin{array}{ccccc}
                                     #1\\#2\\#3\\#4\\#5\end{array}
                              \right)  }
\def\nn               {  \nonumber  }
\def\bracetwo#1#2     {  \left\{ \begin{array}{l} #1 \\ #2 \end{array}
                         \right.  }
\def\bracetwocases#1#2#3#4  {   \left\{ \begin{array}{ll} #1 &
                                 \qquad #2 \\
                                 #3 & \qquad #4 \end{array} \right.  }
\def\bracebegin#1     {  \left\{ \begin{array}{#1}   }
\def\braceend         {  \end{array}\right.   }
\def\del        {  \partial }
\def\half       {  {1\over 2}  }
\def\trace      {  \mbox{Tr}\,  }

\def\abs#1      {  \vert #1 \vert  }

\def\e          { {\rm e}  }
\def\comma          {\, ,}
\def\period         {\, .}
\def\lsim      {\lower .65ex \hbox{\ $\stackrel{<}{\sim}$\ } }
\def\gsim      {\lower .65ex \hbox{\ $\stackrel{>}{\sim}$\ } }

\def\dim       {{\rm dim}\, }
\def\bra#1{{\langle #1 | } }
\def\ket#1{{| #1 \rangle } }
\def\com#1#2{{ \left[#1, #2\right] } }
\def\acom#1#2{{ \left\{ #1,#2\right\} } }
\def\nn               {  \nonumber  }
\newcommand{\nullify}[1]{}
\def\bfall{\boldmath\bf }

\def\Dcom#1#2{\left\{#1, #2\right\}_D}

\def\th{\theta}
\def\alp{{\al'}}

\def\xdot{\dot{x}}
\def\thdot{\dot{\theta}}
\def\etadot{\dot{\eta}}

\def\Stil{\tilde{S}}

\def\etatil{\tilde{\eta}}

\def\xbar{{\bar{x}}}

\def\abar{{\bar{a}}}
\def\bbar{{\bar{b}}}
\def\albar{{\bar{\alpha}}}

\def\mubar{{\bar{\mu}}}
\def\nubar{{\bar{\nu}}}
\def\rhobar{{\bar{\rho}}}
\def\ahat{{\hat{a}}}
\def\bhat{{\hat{b}}}
\def\chat{{\hat{c}}}
\def\dhat{{\hat{d}}}
\def\lhat{{\hat{l}}}
\def\Ahat{{\hat{A}}}

\def\alhat{{\hat{\alpha}}}
\def\ihat{{\hat{i}}}
\def\jhat{{\hat{j}}}

\def\mbJ{\mathbb{J}}
\def\mbP{\mathbb{P}}
\def\mbK{\mathbb{K}}
\def\mbH{\mathbb{H}}
\def\mbD{\mathbb{D}}
\def\mbE{\mathbb{E}}

\def\mbQ{\mathbb{Q}}
\def\mbS{\mathbb{S}}
\def\mbJhat{\hat{\mathbb{J}}}
\def\mbPhat{\hat{\mathbb{P}}}
\def\mbKhat{\hat{\mathbb{K}}}
\def\mbDhat{\hat{\mathbb{D}}}

\def\mbQhat{\hat{\mathbb{Q}}}
\def\mbShat{\hat{\mathbb{S}}}
\def\una{{\underline{a}}}
\def\unb{{\underline{b}}}

\def\ovsqtwo{{1\over \sqrt{2}}}
\def\sqtwo{\sqrt{2}\, }

\def\Dcom#1#2{\left\{#1, #2\right\}_D}

\def\Omlket{\ket{\Omega_l}}
\def\Psiket{\ket{\Psi}}
\def\Psilket{\ket{\Psi_l}}
\def\NS{N_S} 
\def\NStil{N_{\tilde{S}}}

\def\Str{{\rm Str}}

\def\lket{\ket{0,l,0}}

\def\Ap{{A'}}
\def\Bp{{B'}}
\def\Cp{{C'}}
\def\Dp{{D'}}
\def\papertitlepage{\baselineskip 3.5ex \thispagestyle{empty}}
\def\Title#1{\baselineskip 1cm \vspace{1.5cm}\begin{center}
 {\Large\bf #1} \end{center}
\vspace{0.5cm}}
\def\Authors#1{\begin{center} {\it #1} \end{center}}
\def\Abstract{\vspace{1.0cm}\begin{center} {\large\bf Abstract}
           \end{center} \par\bigskip}

\def\Komabanumbertwo#1#2{\hfill \begin{minipage}{4.2cm} UT-Komaba #1
              \parn #2 \end{minipage}}
\renewcommand{\thefootnote}{\fnsymbol{footnote}}
\renewenvironment{thebibliography}{\pagebreak[3]\par\vspace{0.6em}
\begin{flushleft}{\large \bf References}\end{flushleft}
\vspace{-1.0em}

\begin{enumerate}\if@twocolumn\baselineskip=0.6em\itemsep -0.2em
\else\itemsep -0.2em\fi\labelsep 0.1em}{\end{enumerate} }

\begin{document}
\papertitlepage
\vspace*{0cm}
\Komabanumbertwo{09-7}
{December, 2009}
\Title{Exact Quantization of a Superparticle in {\bfall $AdS_5\times S^5$} }
\Authors{{\sc Tetsuo Horigane\footnote[2]{horigane@hep1.c.u-tokyo.ac.jp}
 and Yoichi Kazama\footnote[3]{kazama@hep1.c.u-tokyo.ac.jp}
\\ }
\vskip 3ex
 Institute of Physics, University of Tokyo, \\
 Komaba, Meguro-ku, Tokyo 153-8902 Japan \\
  }
\baselineskip .7cm
\numberwithin{equation}{section}
\numberwithin{figure}{section}
\numberwithin{table}{section}
\parskip=0.9ex
\vspace{1.7cm}
\Abstract
As a step toward deeper understanding of the AdS/CFT correspondence,  exact 
quantization of a Brink-Schwarz superparticle in the $AdS_5\times S^5$ background  with Ramond-Ramond (RR) 
 flux is performed from the first principle in the phase space formulation. 
It includes the construction of the  quantum Noether charges for the $psu(2,2|4)$ superconformal symmetry and by solving the superconformal primary conditions 
 we obtain the complete physical  spectrum of the system with the explicit wave
 functions.  The spectrum  agrees precisely  with the supergravity results,
 including all the Kaluza-Klein excitations. Our method and 
 the result  are expected to shed light on the eventual quantization of  a superstring 
 in this important background. 
\newpage
\baselineskip 3.5ex
\section{Introduction }
\renewcommand{\thefootnote}{\arabic{footnote}}
For more than a decade since its inception, the concept of AdS/CFT\cite{Maldacena:1997re,Gubser:1998bc,Witten:1998qj} has been 
an inexhaustible  source of new developments in both string theory and quantum field theory.  In recent years  it has been applied to such  broad areas as  QCD phenomenology\cite{Sakai:2004cn,Erlich:2005qh,Da Rold:2005zs}, condensed matter 
physics\cite{Hartnoll:2009sz}  and so on that if successful its magical power would be 
even more enhanced. It  is ``magical" since, despite the existence of a pile of impressive 
 evidence, the understanding of the fundamental mechanism of this correspondence is still a difficult  unsolved problem. 

Evidently, the major reason for this difficulty lies in the strong/weak nature of the 
 correspondence. In the prototypical example  of the correspondence between 
 the $N=4$ super-Yang-Mills (SYM) theory in 4 dimensions and the type 
IIB superstring 
in $AdS_5\times S^5$ with RR flux,  which will be the exclusive focus of our attention  in this article,  it is expressed by the well-known  relation 
$g_{YM}^2 N = 4\pi g_s N = R^4/{\alp}^2$, where $R$ is the common radius 
of $AdS_5$ and $S^5$. This succinctly  expresses the fact that  
large  't Hooft coupling on the CFT side corresponds to the weak coupling 
on the string worldsheet and vice versa. To understand the physical meaning  of this relation, one notes that it contains two equalities of different nature. The first equality signifies the familiar open-closed duality, which holds perturbatively. The second equality on the other hand refers only to the closed string side. It can be interpreted as 
 expressing the fact that the metric and the RR 5-form condense in tandem  to produce the $AdS_5 \times S^5$ with common radius $R$. In fact the action density of the metric is given by $\calR/(g_s^2 l_s^8) \sim 1/(g_s^2 l_s^8 R^2)$, where $\calR$ is the scalar curvature and $l_s$ is the string scale, while the action density  of the $N$ units of 5-form flux is  $F_5^2 \sim (N/R^5)^2$, where $R^5$
 is the volume of $S^5$. Equating these two expressions one immediately obtains 
(the square of) the  second equality. This suggests that 
to understand the AdS/CFT correspondence dynamically one would have to sum over
 the infinite number of open string loop diagrams attached to a stack of D-branes 
and interpret it  from the closed string channel as condensation of the metric and 
the RR 5-form which warps  the spacetime. In other words, it tantamounts to 
showing rigorously  that the D-branes are what we believe they are. 
Attempts along this line  have been made 
recently\cite{Kruczenski:2006ti, Kruczenski:2007jg}\cite{Kawai:2007ek, Kawai:2007eg}, but a precise tractable formulation appears  to be hard at the moment.  

In short of the fundamental dynamical understanding, the next best thing is to 
demonstrate  that the symmetry structure, the spectrum,  and the correlation functions of  basic physical quantities match exatly on both sides of the correspondence.
 In  mathematical sense, this would constitute  
a proof of the equivalence of two theories. 
This  is the spirit of the celebrated 
Gubser-Klebanov-Polyakov-Witten (GKP-W) relation\cite{Gubser:1998bc,Witten:1998qj} and it has been 
quite  successful  for the  $\half$ BPS quantities, largely  because 
the supergravity approximation  can be used  on the string side.  

To go beyond this
  approximation, the main difficulty resides in the extension of supergravity 
 to incorporate the stringy excitations. 
The most  direct way would be  to  construct  a closed superstring field theory 
in the $AdS_5\times S^5$  background, but it  appears to be beyond reach   at the present time.  A more practical  approach is to develop a worldsheet first quantized  formalism 
and  compute the correlation functions by constructing  appropriate vertex operators  anchored at  the points on the boundary of $AdS$ spacetime. Research in this direction was initiated in \cite{Metsaev:1998it} using  the Green-Schwarz formalism\cite{Green:1983wt, Green:1983sg} 
and subsequently in the pure spinor formalism\cite{Berkovits:2000fe}.  Since then 
  numerous investigations were made but most of them are classical 
 or semi-classical and a full fledged quantization of 
a superstring ({\it i.e.} to all orders in $\al'$) in $AdS_5 \times S^5$ background has not  been achieved. Consequently, the precise spectrum of the theory is not yet known. 
For  recent reviews, readers are referred to \cite{Tseytlin:2009fw,Bedoya:2009np,Arutyunov:2009ga}and references therein. 

As a matter of fact, even a superparticle\cite{Brink:1981nb}, 
which represents  the zero mode of the superstring, 
 has not  been systematically quantized from the first principle 
in this curved background.  We should note, however,  that in a pioneering work
\cite{Metsaev:1999gz} Metsaev wrote down  a quadratic action for 
a light-cone superfield, which was invariant under a set of 
 $psu(2,2|4)$ generators made out of  the coordinates and the momenta 
of a superparticle.   Although the method was not systematic, 
 this was equivalent to quantization of a superparticle.  Concerning  the spectrum 
of this system, some analysis of the AdS ``mass" operator was performed 
 but  the AdS energy  spectrum was not obtained.  In subsequent developments\cite{Metsaev:1999kb, Metsaev:2002vr}, 
 the AdS energy was worked out for some subset of the states and was shown to 
agree with that of the corresponding supergravity fields. 
Also,  advancements were made for the formalism itself,  as a part of the  
formulation of the superstring. Classical action for a superstring in the 
light-cone gauge was derived explicitly 
based  on the supercoset formalism in \cite{Metsaev:2000yf} 
and the construction of the generators of $psu(2,2|4)$ was made 
more systematic in \cite{Metsaev:2000yu}. Nevertheless, these developments 
were purely classical. 

In this article, we will be able to make substantial progress on the understanding of the quantum aspects  of a Brink-Schwarz 
superparticle in $AdS_5 \times S^5$ with RR flux. 
It consists of  (i) an exact systematic  quantization from the first principle,  including 
 the derivation of the quantum Noether charges for the $psu(2,2|4)$
 (superconformal)   symmetry,  and (ii) complete  solution of the spectrum of the theory with the explicit wave  functions for the superconformal primaries.  This is achieved in the physical light-cone gauge in the phase space formulation.  The spectrum agrees precisely with 
the supergravity results\cite{Kim:1985ez} \cite{Gunaydin:1984fk}, including 
all the Kaluza-Klein excitations. As a superparticle constitutes  the zero-mode part
 of a superstring, our method and the result should shed light on the eventual 
quantization of a  superstring in this important curved background. 

We will now give the outline of our work, which at the same time serves to indicate
 the organization of the rest of this article.  We will begin by describing, in section 2,
 the classical phase space formulation of a superparticle 
 in the $AdS_5\times S^5$ background. 
More specifically, after recalling the $psu(2,2|4)$ symmetry algebra in section 2.1, 
 we will review, in section 2.2,  the supercoset method of constructing the invariant classical action in the ``light-cone gauge", first performed in \cite{Metsaev:2000yf}. Then we will develop the phase space formulation based on such an action in section 2.3. We will develop  a powerful  method of finding the Dirac  brackets for the fundamental physical variables from the gauge-fixed action and  find appropriate combinations which satisfy the canonical form of the bracket relations. 
The section 3 will be devoted to the construction of the quantum Noether charges 
 for the $psu(2,2|4)$ superconformal symmetry.  We will first 
compute the Noether charges at the classical level in terms of the phase space 
 variables and then quantize them by performing  appropriate normal-ordering. 
All the quantum charges are explicitly obtained, which will be important in solving 
the system completely. 
In section 4, which is the main part of this article, we  will give the complete
 solutions for the superconformal primary states of the system and show that the 
spectrum  precisely agrees  with the supergravity results.  
  In preparation for the solution, we  will first discuss, in 
section 4.1, the  two  choices of the scheme  of the representation of the superconformal algebra, which will be called the dilatation (D) scheme  and 
the energy (E) scheme. Then, 
after presenting the superconformal primary conditions in section 4.2, 
we will analyze the allowed highest weight unitary representations for the
 $su(4)$ sector in section 4.3.  Finally in section 4.4, we will solve the 
superconformal primary conditions  to obtain the  wave functions 
explicitly and show that they enjoy expected properties. 
The section 5 is devoted to discussions and future perspectives. 
Several appendices are provided to display  some further details. 
\section{Phase space formulation of a classical superparticle \\ in {\bfall $AdS_5\times S^5$}
with RR flux }
We begin  by  describing  the phase space formulation of a superparticle 
 in $AdS_5 \times S^5$ background with RR flux at the classical level.  
We will adopt  the  the Brink-Green-Schwarz formulation\cite{Brink:1981nb}\cite{Green:1983wt, Green:1983sg}  and 
 basically follow the light-cone-gauge treatment of  Metsaev and Tseytlin
\cite{Metsaev:1998it, Metsaev:2000yf} for a  string in the above  background. 
 Upon dropping the dependence
 on $\sigma$, the coordinate along the string, 
we can specialize to the case of a  particle. 
Therefore  this section is mostly a review,  except  that a
 new  important  observation  will be made in the subsection 2.3 
concerning the systematic computation of the Dirac bracket. 
\subsection{{\bfall $psu(2,2|4)$} algebra in the light-cone basis} 
The most efficient way to construct the (Brink-)Green-Schwarz action
 for a string (and a particle)  in $AdS_5\times S^5$ background with 
RR flux is to make use of the supercoset method\cite{Henneaux:1984mh}\cite{Metsaev:1998it}  based on the global 
symmetry group $PSU(2,2|4)$, the bosonic part of which is 
 $SO(4,2)\times SO(6)$.  Indeed, it is well-known  that $AdS_5\times S^5$ 
 can be represented as the coset 
\begin{align}
AdS_5\times S^5 &\simeq {SO(4,2) \times SO(6) \over SO(4,1) \times SO(5) }
\period 
\end{align}
Therefore we must first discuss the generators of $PSU(2,2|4)$, which form 
 the Lie superalgebra $psu(2,2|4)$. 

The even part of $psu(2,2|4)$ consists of   $so(4,2)$ and $so(6)$. 
$so(4,2)$ can be regarded as acting  on the six-dimentional flat space with coordinates 
$X^A =(X^{-1}, X^0, X^1,X^2,X^3,X^4)$ and  the signature $(-, -,+,+,+,+)$. 
 Its generators, to be denoted by   $T^{AB}$, satisfy the commutation relations 
\begin{align}
\com{T^{AB}}{T^{CD}} &= \eta^{BC} T^{AD} -\eta^{AC}T^{BD}
-\eta^{BD}T^{AC} + \eta^{AD}T^{BC}  \period 
\end{align}
We adopt the convention  that  $T^{AB}$'s  are anti-hermitian. 
$T^{AB}$ can be decomposed with respect to the $so(4,1)$ subalgebra, which will be
denoted as
\begin{align}
T^{AB} &= (T^{\ahat\bhat}, T^\ahat) \period \label{so41dec}
\end{align}
Here  $T^\ahat \equiv T^{\ahat,-1}$ and $\ahat = 0 \sim 4$. 
In the context of AdS/CFT,  it will be useful to regard $SO(4,2)$ as the conformal group 
 in four dimensions.  From this point of view, it is natural to introduce the 
 ``conforaml basis"  generators as\footnote{
Our definitions of $D, P^a$ and $K^a$ differ  slightly from the ones 
 used  in \cite{Metsaev:2000yf}. In particular we take  $D$ to be opposite
 in sign because we prefer to have the momentum $P^a$ to carry the 
dimension $+1$.   }
\begin{align}
P^a &\equiv {1 \over \sqtwo} ( T^a -T^{a4} )   \comma \qquad 
 K^a  \equiv  -{1\over \sqtwo }  (T^a + T^{a4})   \comma \label{defPK}\\
D &\equiv T^4 \comma \qquad J^{ab} \equiv  T^{ab} \comma 
\end{align}
where $P^a, K^a, D, J^{ab}$ are the generators of translation, the special 
 conformal transformation, the dilatation and the Lorentz rotations respectively and 
 the  ``Lorentz index" $a$ runs over the range $0 \sim 3$. 
They satisfy the commutation relations
\begin{align}
\com{J^{ab}}{P^c} &=\eta^{bc} P^a -\eta^{ac}P^b \comma \\
\com{J^{ab}}{K^c} &=\eta^{bc} K^a -\eta^{ac}K^b  \comma \\
\com{P^a}{K^b} &= -\eta^{ab}D -J^{ab} \comma \\
\com{D}{P^a} &= P^a \comma \qquad 
\com{D}{K^a} = -K^a \comma \\
\com{J^{ab}}{J^{cd}} &= \eta^{bc}J^{ad} -\eta^{ac}
J^{bd} -\eta^{bd}J^{ac}+\eta^{ad}J^{bc}  \period
\end{align}
In relation to the $\kappa$-symmetry gauge fixing, to be discussed later, 
we will often use the 
``light-cone basis" (in the sense of four dimensions).  For the basic coordinates 
of the four dimensional space 
 the light-cone components  are defined as
\begin{align}
x^{\pm} &= \frac{1}{\sqrt{2}}( x^3 \pm x^0 ) \comma \qquad 
x = \frac{1}{\sqrt{2}}( x^1 + i x^2) \comma \qquad 
\bar{x} = \frac{1}{\sqrt{2}}( x^1 - i x^2) \period
\end{align}
In other words,  the metric in this basis has  non-vanishing components 
$\eta^{+-}=\eta^{-+}=1$,
$\eta^{x\bar{x}}=\eta^{\bar{x}x}=1$.  Accordingly,  the generators of $so(4,2)$ in this basis will be taken  as 
\begin{align}
P^\pm,  \hspace{4pt}    P^x  ,  \hspace{4pt} 
P^{\bar x}, \hspace{4pt}    K^\pm ,  \hspace{4pt}   K^x  ,   \hspace{4pt}
 K^{\bar x}, \hspace{4pt} D, \hspace{4pt} 
 J^{+-} , \hspace{4pt}  J^{\pm x} , \hspace{4pt}
  J^{\pm\bar x} ,  \hspace{4pt}  
J^{x\bar x} \period 
 \end{align}
Further we will employ the following simplified notations 
\begin{equation}
 P\equiv P^x ,  \hspace{4pt}
\bar{P} \equiv P^{\bar x} ,  \hspace{4pt} K\equiv K^x ,
 \hspace{4pt}  \bar{K} \equiv K^{\bar x} .
\end{equation}
From  these definitions it is straightforward to write down  the commutation relations 
for  the generators in the light-cone basis. 

Next consider the remaining bosonic subalgebra $so(6)$. This will be interpreted as 
$su(4)$ since the fermionic generators of $psu(2,2|4)$ transform under the fundamental and anti-fundamental representations of $su(4)$. The traceless $su(4)$ generators $J^i{}_j$ $(i,j=1\sim 4)$   satisfy the algebra
\begin{align}
[J^i{}_j,J^k{}_n]=\delta^k_jJ^i{}_n
-\delta^i_n J^k{}_j \period  \label{su4com}
\end{align}

Now we come to the odd part of the $psu(2,2|4)$ algebra. It consists of 
 32 supercharges   $Q^{\pm i}$, $Q_i^\pm$, $S^{\pm i}$, $S_i^\pm$ $(i=1\sim 4)$, which transform, as said above,  under $su(4)$ as 
\begin{align}
\com{J^i{}_j}{ Q^\pm_k} &= -\delta^i_k Q^\pm_j + {1\over 4} \delta^i_j Q^\pm_k \comma \qquad 
\com{J^i{}_j}{ Q^{\pm k}} = \delta^k_j Q^{\pm i} - {1\over 4} \delta^i_j 
Q^{\pm k } \comma 
\end{align}
and similarly for the $S$ supercharges.  The superscripts $\pm$ on $Q$ and $S$ 
generators indicate  their charge with respect to  the generator $J^{+-}$ ({\it i.e.} 
the boost along the 3-direction),  as the following commutation relations show:
\begin{align}
& [J^{+-},Q^{\pm i}]
=\pm\frac{1}{2}Q^{\pm i}\comma  \qquad 
[J^{+-},Q^\pm_i]
=\pm\frac{1}{2}Q^\pm_i \comma \qquad \\
& [J^{+-},S^{\pm i}]
=\pm\frac{1}{2}S^{\pm i}\comma \qquad 
[J^{+-},S^\pm_i]
=\pm\frac{1}{2}S^\pm_i \period
\end{align}
We note that,  together  with the commutation relations for the bosonic generators already  given,  the values of the $J^{+-}$-charge 
 for all the generators are in the finite range  $[-1 , +1]$. 
 This fact will  play an  important role in the gauge fixing later. 

The fermionic generators  also carry  charges with respect to 
 the generators  $D$  and $J^{x\bar x}$. The charge assignment
is  expressed through the following commutation relations:
\begin{align}
 &\qquad  [D,Q^{\pm i}]=\frac{1}{2}Q^{\pm i}\comma  \qquad 
[D,Q^\pm_i]=\frac{1}{2}Q^\pm_i \comma \\
 &\qquad  
[D,S^{\pm i}]=-\frac{1}{2}S^{\pm i} \comma \qquad 
[D,S^\pm_i]=-\frac{1}{2}S^\pm_i \comma \\
&\qquad [J^{x\bar{x}},Q^{\pm i}]=\pm\frac{1}{2}Q^{\pm i} \comma 
\qquad 
[J^{x\bar{x}},Q^\pm_i]=\mp\frac{1}{2}Q^\pm_i \comma  \\
& \qquad [J^{x\bar{x}},S^\pm_i]=\pm\frac{1}{2}S^\pm_i \comma \qquad 
[J^{x\bar{x}},S^{\pm i}]=\mp\frac{1}{2}S^{\pm i} \period
\end{align}

The transformation properties of the supercharges under the four dimensional 
 Poincar\'e generators are as follows.  Under the Lorentz rotations they 
 transform as 
\begin{align}
&[J^{+x}, Q^{-i}]=Q^{+i},
&&[J^{+\bar{x}}, S^{-i}]=S^{+i},
&&[J^{-\bar{x}}, Q^{+i}]=-Q^{-i},
&&[J^{-x}, S^{+i}]=-S^{-i}, \\
&[J^{+\bar{x}}, Q^-_i]=Q^+_i,
&&[J^{+x}, S^-_i]=S^+_i,
&&[J^{-x}, Q^+_i]=-Q^-_i,
&&[J^{-\bar{x}}, S^+_i]=-S^-_i \comma 
\end{align}
while the commutation relations with the translation and the conformal boost generators take the form 
\begin{align}
&[P^\pm, S^{\mp i}]=  i Q^{\pm i},
&&[\bar{P}, S^{-i}]=  i Q^{-i},
&&[ P, S^{+i}]= -  iQ^{+i},
\\
&[P^\pm, S^\mp_i]=- i Q^\pm_i,
&&[P, S^-_i]=-  i Q^-_i,
&&[\bar{P}, S^+_i]= i Q^+_i,
\\
&[K^\pm, Q^{\mp i}]= i S^{\pm i},
&&[K, Q^{-i}]= i S^{-i},
&&[\bar{K}, Q^{+i}]=- i S^{+i}, \\
&[K^\pm, Q^{\mp}_i]= -  i S^{\pm}_i,
&&[\bar{K}, Q^-_i]= - iS^-_i,
&&[ K, Q^+_i]=  i S^+_i.
\end{align}
Finally, the anticommutation relations between the supercharges are given by 
\begin{align}
&\{Q^{\pm i},Q_j^\pm\}=\mp  i  P^\pm\delta^i_j, \hspace{6pt}
&&\{Q^{+i},Q_j^-\}=-  iP\delta^i_j, \hspace{6pt}
&&\{Q^+_i, Q^{-j} \} = -i \bar{P}\delta^i_j, \\
&\{S^{\pm i},S_j^\pm\}=\pm  i K^\pm\delta^i_j, \hspace{6pt}
&&\{S^{-i},S_j^+\}=i K\delta^i_j, \hspace{6pt}
&&\{ S^-_i , S^{+j} \} = i \bar{K}\delta^i_j,
\end{align}
\begin{align}
&\{Q^{+i},S^+_j\}=-J^{+x}\delta^i_j \comma \qquad 
\{Q^+_i, S^{+j} \} = J^{+\bar{x}}\delta^j_i \comma \\
&\{Q^{-i},S^-_j\}=-J^{-\bar{x}}\delta^i_j \comma \qquad 
\{ Q^-_i, S^{-j}\} = J^{-x}\delta_i^j \comma 
\end{align}
\begin{align}
& \{Q^{\pm i},S^\mp_j\}=\frac{1}{2}(J^{+-}+J^{x\bar{x}}\pm D)\delta_j^i
\ \mp \ J^i{}_j \comma  \\
& \{Q^\pm_ i,S^{\mp j}\}=\frac{1}{2}(-J^{+-}+J^{x\bar{x}}\mp D)\delta_i^j \ \mp \ J^j{}_i \period 
\end{align}
Hermiticity properties of the generators are such as to be  consistent with 
 the $psu(2,2|4)$ algebra. Explicitly, 
\begin{align}
& (P^{\pm})^\dagger=-P^\pm\comma 
\quad
P^\dagger=-\bar{P}\comma 
\quad
(K^\pm)^\dagger=-K^\pm\comma 
\quad
K^\dagger=-\bar{K}\comma   \\
&
(J^{\pm x})^\dagger=-J^{\pm \bar{x}} \comma 
\quad
(J^{+-})^\dagger=-J^{+-} \comma 
\quad
(J^{x\bar{x}})^\dagger=J^{x\bar{x}}\comma 
\quad
D^\dagger=-D \comma \\
\quad
&(J^i{}_j)^\dagger=J^j{}_i \comma  \quad 
(Q^{\pm i})^\dagger=Q^{\pm}_i \comma 
\quad     
(S^{\pm i})^\dagger=S^{\pm}_i \period \label{hermite}
\end{align}
This completes the description of the $psu(2,2|4)$ algebra in the light-cone basis. 
\subsection{Supercoset construction}
We are now ready to construct the action by the supercoset method. 
The supercoset of interest is $\calK = \calG/\calH$ where 
\begin{align}
\calG = PSU(2,2|4)\comma \qquad \calH = SO(4,1) \times SO(5)  \period 
\end{align}
We follow \cite{Metsaev:2000yf} and take the representative element $G$ of 
$\calK$ in the form  
\begin{align}
G &= g_x g_{\theta}  g_\eta  g_\phi g_y \comma \\
g_x &= \exp (x^a P_a) \comma \\
g_{\theta} &= \exp \left(   \th^{-i} Q^+_i 
 + \th_i^- Q^{+i} + \th^{+i} Q^-_i + \th^+_i Q^{-i} \right)   \comma \\
g_\eta&=\exp\left(  \eta^{-i} S^+_i 
 + \eta_i^- S^{+i} + \eta^{+i} S^-_i + \eta^+_i S^{-i} \right)  \comma \\
g_\phi&= \exp (\phi D) \comma \\
g_y &= \exp( y^i{}_j J^j{}_i  )   \comma \qquad y^i{}_j = {i \over 2} (\ga^{A'})^i{}_j  y^{A'} \comma \quad A'=5 \sim 9  
\period \label{eqy} 
\end{align}
The variables $(x^a, \phi)$ describe  the $AdS_5$ part, while  $y^{A'}$ are the 
coordinates of $S^5$. 
 The fermionic 
 part of the coset is parametrized by the grassmann variables 
  $\th^{\pm i}, \th^\pm_i, \eta^{\pm,i}, \eta^\pm_i$, with the 
 conjugation property $\theta^\pm_i = (\theta^{\pm i})^\dagger$, 
$\eta^\pm_i = (\eta^{\pm i})^\dagger$.

Perhaps a clarifying remark should be made on the choice of the coset parametrization, 
especially  the part which is supposed  to parametrize  the $AdS_5$, {\it i.e.} 
 $g_{x,\phi} = \exp(x^a P_a) \exp(\phi D)$.  At first sight one might 
 worry  since  the generators of $SO(4,2)$ 
which do not appear in $g_{x,\phi}$  are $\{K^a, J^{ab}\}$ and 
they generate a  group  isomorphic to  the Poincar\'e group 
 in four dimensions (with $K^a$ playing the role of the translation operators), 
which is not $SO(4,1)$ but rather its contraction limit. 
Thus it would seem more legitimate to take $\exp(x^\ahat T_\ahat)$
 as the coset representative since we already saw in (\ref{so41dec}) that 
 a natural  decomposition of the generators of $SO(4,2)$ into $SO(4,1)$ and the coset part is given by   $(T^{\ahat \bhat}, T^\ahat)$.  Actually, the choice of $g_{x,\phi}$ 
is perfectly legitimate. The reason  is that 
an arbitrary element of $SO(4,2)$ can be shown to be represented  in the form 
$g_{x,\phi} h$, where $h= \exp(y_{\ahat\bhat} T^{\ahat\bhat})
\in SO(4,1)$. (In fact we can use any  embedding of $SO(4,1)$ in $SO(4,2)$ for 
this purpose. )  All  we have to make sure  is that because 
 $(P_a, D)$ do not coincide with  the coset directions $T^\ahat$ 
we must  project out the motion along the coset manifold properly, 
 as we will  explain shortly.

As is well-known, the basic building  block  of the supercoset method is 
 the Maurer-Cartan (MC) 1-form  $J= G^{-1} dG$,  which is   invariant under 
 the left action of $PSU(2,2|4)$. As it takes its  value in $psu(2,2|4)$, it can be 
 expanded  as
\begin{align}
J &= G^{-1}dG = L^a_P P_a + L^a_K K_a 
+ L_DD ++ \frac{1}{2}L^{ab}_JJ^{ab}
 + L^j{}_{i}J^i{}_j \nn\\
& + L^{+k}_QQ^-_k +  L^+_{Qk}Q^{-k} +   L^{-k}_QQ^+_k +  L^-_{Qk}Q^{+k} \notag \\
 & +  L^{+k}_SS^-_k +  L^+_{Sk}S^{-k} +  L^{-k}_SS^+_k +  L^-_{Sk}S^{+k} 
\period \label{MCform}
\end{align}
In contrast to  the case of the flat space time, all the generators, not just the coset generators,  appear on the right hand side.  This means that $G^{-1} dG$ as a whole 
 describes the motion in the entire group space. 
What we really want is the motion along the bosonic 
 coset, namley $AdS_5\times S^5$. To extract this out, we need the orthogonal 
decomposition of  the coset part and the rest. 
This is achieved by the use of the  invariant bilinear form,  commonly 
 called the ``supertrace".  It is given by\cite{Berkovits:1999zq}
\begin{align}
\Str(T^\ahat T^\bhat) &=\eta^{\ahat \bhat}  \comma \qquad 
\Str(T^{\ahat \bhat} T^{\chat \dhat})  = \eta^{\bhat\chat}\eta^{\ahat\dhat}
-\eta^{\ahat\chat}\eta^{\bhat\dhat}  \comma \label{strso42}\\
\Str(J^i{}_j J^k{}_n) &= 
-\delta^i_n \delta^k_j + {1\over 4} \delta^i_j \delta^k_n 
\comma \label{strsu4} \\
\Str(Q^{\pm i} S^{\mp}_j) &= \Str(Q^{\mp}_i S^{\pm j}) = \pm \delta^i_j
  \comma \\
  \mbox{Rest} &=0   \period
\end{align}
Concerning the $SO(4,2)$ sector, 
we see from (\ref{strso42}) that 
  $T^\ahat= (T^a, T^4)$ are the desired coset 
generators which are  orthogonal to the $so(4,1)$ generators $T^{\ahat\bhat}$. 
As for the $SO(6)$ sector, we need to first convert the  $SU(4)$ generators $J^i{}_j$ to $SO(6)$ generators and then decompose them 
 into the  $SO(6)/SO(5)$ part and the $SO(5)$ part. This is achieved 
with  the aid  of the $4\times 4$ $\gamma$-matrices $\ga^{A'}$ of $SO(5)$ 
and its antisymmetrized products $\ga^{\Ap\Bp}$:
If we define 
\begin{align}
J^\Ap &\equiv  -{i \over 2} J^i{}_j (\ga^\Ap)^j{}_i \comma \qquad 
J^{\Ap\Bp}  \equiv  -\half J^i{}_j (\ga^{\Ap\Bp})^j{}_i \comma 
\label{so6gen}
\end{align}
one can  check that they together form the $so(6)$ algebra, 
 where $J^{\Ap\Bp}$ generate the $so(5)$ and $J^\Ap$ represent the 
coset generators. Furthremore, from (\ref{strsu4}) one can easily obtain
\begin{align}
\Str (J^\Ap J^\Bp) &= \delta^{\Ap\Bp} \comma \qquad 
\Str(J^{\Ap\Bp} J_{\Cp\Dp} ) =  \delta^{\Ap\Bp}_{\Cp\Dp} 
\comma \qquad \Str(J^\Ap J^{\Bp\Cp}) =0 
\comma 
\end{align}
 showing that the $J^\Ap$ form  the desired orthogonal basis of  the coset $SO(6)/SO(5)$. 
Therefore, we should rewrite the MC 1-form (\ref{MCform}) as 
$J = J_B + \mbox{rest}$ and extract  the bosonic coset part  $J_B$ defined  by 
\begin{align}
J_B &\equiv  J_B^\ahat T_\ahat + J_B^\Ap J_\Ap \period
\end{align}
From the definitions (\ref{defPK}) and (\ref{so6gen}) we  find 
\begin{align}
J_B^\ahat &=(J^a_B, J^4_B) = \left( \ovsqtwo (L^a_P -L^a_K), L_D\right) 
\comma \qquad  J_B^\Ap = {i \over 2} (\ga^\Ap)^i{}_j L^j{}_i \period 
\end{align}
Then  the Lagrangian of a superparticle in $AdS_5\times S^5$ is given by 
\begin{align}
L &= {1\over 2e} \Str (J_B J_B) = {1\over 2e} 
\left( \eta_{\ahat\bhat} J^\ahat_B J^\bhat_B + \delta_{\Ap\Bp}
 J^\Ap_B J^\Bp_B \right) \comma \label{genaction}
\end{align}
where $e$ is the einbein and  we have used 
 the same symbol $J_B$ to mean  the coefficient of $d\tau$
 in the 1-form $J_B$, where $\tau$ is the parameter along the worldline. Note that for a superparticle the Wess-Zumino term 
which is crucial for the  $\kappa$-invariance in the superstring case
  vanishes since it  contains a derivative with respect to $\sigma$. 
Indeed,  the  action above already possesses the desired $\kappa$ symmetry. 

Although the Lagrangian above has the virture of being 
 manifestly invariant under the $psu(2,2|4)$ symmetry, 
 it cannot be computed explicitly.  
The reason is that the MC 1-form $G^{-1}dG$ can contain up to 32 powers 
 of fermionic coordinates and it is practically impossible to compute it in closed form. 
This problem can be solved by imposing  judicious  gauge conditions. 
A convenient set of 16 conditions we adopt are the so-called 
  semi-light-cone gauge conditions given by 
\begin{align}
\theta^{+i} &= \theta^+_i = \eta^{+i} = \eta^+_i = 0 \comma 
\end{align}
which will often be denoted simply as $\Theta^+_I=0$. 
This means that only the supercharges with the $J^{+-}$ charge $+\half$ are kept
 in the coset representative $G$. Consequently, $g_\theta$ and $g_\eta$ are 
 reduced to  
\begin{align}
 g_{\theta} &= \exp \left(   \th^{i} Q^+_i 
 + \th_i Q^{+i} \right)   \comma \qquad 
g_\eta =\exp\left(  \eta^{i} S^+_i 
 + \eta_i S^{+i}  \right)   \comma \\
\theta^i &\equiv \theta^{-i} \comma \quad \theta_i \equiv \theta^-_i 
\comma \quad \eta^i \equiv \eta^{-i} \comma \quad \eta_i 
 \equiv \eta^-_i \period 
\end{align}
Here and hereafter, we suppress   the 
 superscript ``$-$" for the  the remaining ferminonic coordinates 
 for simplicity. 

In this gauge, because the maximum value of the $J^{+-}$ charge for the $psu(2,2|4)$ generators 
 is $+1$,  the expansion of the MC 1-form $J$ in powers of $\theta$ and $\eta$ 
 terminates in a few steps and one obtains simple explicit expressions for the 
 components of $J$. The ones needed to construct the action take the form
\cite{Metsaev:2000yf}
\begin{align}
L_{P^+} &= e^{-\phi} dx^+ \comma \qquad 
  L_{P^-}=e^{-\phi} \left(dx^-
- \frac { i}{2}\tilde{\theta}^i\tilde{d\theta}_i
-\frac{ i}{2}\tilde{\theta}_i \tilde{d{\theta^i}}\right) \comma \label{MCP} \\
L_{P^x} &=e^{-\phi} dx\comma  \qquad 
 L_{P^{\bar{x}}} =e^{-\phi} d\bar{x}, \\
 L_{K^+} &= L_{K^x} = L_{K^{\bar{x}}} = 0 \comma \qquad 
L_{K^-}=e^{\phi}\left(\frac{1}{4}(\tilde{\eta}^2)^2 dx^+
+\frac{ i }{2}\tilde{\eta}^i \tilde{d\eta}_i
+\frac{i }{2}\tilde{\eta}_i \tilde{d{\eta^i}}\right) \comma \\
L_D&=d\phi\comma \qquad 
L^i{}_j= (dUU^{-1})^i{}_j+ i \left( \tilde{\eta}^i\tilde{\eta}_j
-\frac{1}{4}\tilde{\eta}^2\delta_j^i \right) dx^+ \period \label{MCD}
\end{align}
Here the matrix $U$ is given by 
$U = \exp((i/2) y^\Ap \ga^\Ap)$ and 
 the tilded fermionic variables are defined as 
\begin{equation}
\tilde{\theta}^i \equiv  U^i{}_j \theta^j \comma \quad 
\tilde{\theta}_i \equiv  \theta_j (U^{-1})^j{}_i \comma \quad 
\tilde{d{\theta^i}}\equiv U^i{}_j d\theta^j \comma \quad 
\tilde{d\theta}_i \equiv  d\theta_j (U^{-1})^j{}_i,
\end{equation}
and similarly for for $\tilde{\eta}$'s. (More explicit form of $U$ is displayed  in 
 the Appendix A.)
Substituting these expressions into (\ref{genaction}) 
the action is easily obtained as 
\begin{align}
S &= \int d\tau  {1\over2e} \Biggl( e^{-2\phi} \left( \xdot^+ \xdot^- + \xdot \dot{\xbar}  + e^{2\phi} (\dot{\phi})^2  \right)  + (e_0^\Ap)^2 \nn\\
&\qquad -{i \over 2} \xdot^
+ \left[  e^{-2\phi} (\th^i \thdot_i + \th_i \thdot^i) 
+\eta^i \etadot_i + \eta_i \etadot^i  -2i e_0^{A'}  \etatil_i (\ga^{A'})^i{}_j \etatil^j \right]\nn\\
& \qquad - {1\over 4} (\xdot^+)^2 \left[ (\eta^2)^2 - (\etatil_i (\ga^{A'})^i{}_j \etatil^j)^2 \right] 
\Biggr) \comma 
\end{align}
where $e_0^{A'} = -{i \over 2} \trace
(\ga^{A'} \dot{U} U^{-1} ) $.  
Note that if we define  a variable $z$ by
$z \equiv e^\phi$,  the first three  terms can be rewritten as 
\begin{align}
{1\over z^2} \left[\half  \left( {dx^a \over d\tau}\right)^2 + \left( {dz \over d\tau} 
\right)^2
\right] \period 
\end{align}
This shows  that the present  parametrization of the coset 
 corresponds  to  the familiar Poincar\'e coordinates for the $AdS_5$ part, 
up to a trivial scaling. 
\subsection{Classical phase space formulation }
In the preceding subsection we reviewed the construction of  the gauge-fixed 
action for a superparticle in 
the $AdS_5\times S^5$ background in the configuration space.  With the use of the  light-cone gauge  the form of the action has been simplified substantially. 
 Nevertheless  it is still quite  non-linear and  it is difficult to obtain the general solutions of the equations of motion, which are needed for the canonical quantization
 procedure. 

In such a situation  the phase space formulation can be  quite powerful. 
In particular,  when the generator of the dynamics  is contained in the 
symmetry algebra, we may first perform the quantization at equal time without 
solving the dynamical equations of motion and then generate the dynamics 
 algebraically by a  member of the algebra\footnote{ This feature was emphasized and 
 utilized in \cite{Kazama:2008as}  in the analysis of  the superstring in the plane-wave background in the semi-light-cone conformal gauge.}.   This applies to the present case, 
 where 
the generators relevant for the dynamics, namely  the AdS energy operator $E=
-i(P^0-K^0)/\sqrt{2}$ and 
 the light-cone Hamiltonian operator $-P^-$,  are in the $psu(2,2|4)$ algebra. 
For this reason  we will develop the phase space formulation for our system 
in this subsection first  at the classical level. In the next section we will perform the quantization and 
construct the quantum Noether charges which generate  the $psu(2,2|4)$ algebra. 

Although the general procedure for the phase space formulation is a textbook matter,  it is not so easy to execute it in the present case because we do not have the 
 explicit form of the un-gauge-fixed action:  All  we have is the  action on the 
gauge slice $\Theta^+_I=0$.
 In fact we face a trouble right from 
the beginning  since obviously the momenta $(P_{\theta^+}, P_{\eta^+})$ 
conjugate to these variables cannot be computed.  To avoid this problem, 
 one  needs to compute  the action  at least up to  first order in 
$\dot{\Theta}^+_I$, where $\Theta^+_I$ denotes $(\theta^+, \eta^+)$ collectively. 
Suppose we have obtained  such an action with additional efforts. Then we can 
 compute the momenta, define the Poisson brackets for  the basic phase 
 space variables, and find all the constraints  \`a la Dirac. 
Let us focus among them on the first class  fermionic constraints
expressing  the $\kappa$ symmetry  and denote them by $\Phi_J=0$. 
To fix the gauge by  the conditions  $\Theta^+_I=0$ and compute the 
Dirac bracket, we need to know the knowledge of the Poisson brackets 
among the constraints, including the gauge fixing conditions. Since 
$\Phi_J$ can contain the fermionic momenta $P_{\Theta^+} = \del L/\del \dot{\Theta}^+$ as well as $\Theta^+ $ variables, 
this computation actually requires the knowledge of the action to 
order $\Theta^+$ and $ \Theta^+ \dot{\Theta}^+$  as well. 

Summarizing, to execute the usual   procedure for the phase space 
 formulation of our  system of interest, we need to know 
 the action not only on the gauge-slice but also slightly away from it, 
 to order $\Theta^+, \dot{\Theta}^+$ and $\Theta^+ \dot{\Theta}^+$. 
Because of this reason, logically satisfactory  derivation of the Dirac brackets 
has not been performed in the past based on the gauge-fixed action\footnote{
In \cite{Metsaev:2000yu} the Dirac brackets were derived 
 by applying the usual Dirac's method  directly to the light-cone-gauge-fixed action. 
Although this turned out to yield the correct brackets for this system, 
it is not guranteed to be a legitimate procedure in general. 
The reason is as follows: 
To compute the Dirac bracket one needs the  knowledge  of the  matrix $D$ 
formed by the Poisson brackets among the constraints.  Let us  denote 
by $D_g$ and $D_{ung}$ such matrices obtained from the gauge-fixed 
 and the un-gauge-fixed actions respectively. Clearly $D_g$ is a submatrix 
 of $D_{ung}$ because there are less constraints for the gauge-fixed theory. 
Now  in order for the procedure starting 
 from the gauge-fixed action to yield the correct Dirac bracket for the 
 physical variables, the inverse 
$D_g^{-1}$ must  be realized as a block submatrix in  $D_{ung}^{-1}$. 
This however is not necessarily true and has to be checked. 
}. 

We now make an important observation that,  despite the apparent lack of 
 the necessary  information,  there is in fact a systematic way to compute the Dirac 
 brackets using only the knowledge of the gauge-fixed form of $G$ and  $G^{-1}dG$, with a small assumption which  will be a posteriori justified. 
The basic idea is  that, instead of 
computing  the Dirac bracket directly, we will derive the general formula  for the 
 Lagrange bracket, which  is the inverse of the Dirac bracket.  Then we recognize 
 that for the ``physical"  variables, {\it i.e.} the variables other than $\Theta^+$, 
 the formula  for the Lagrange bracket does not contain the derivatives with respect  to $\Theta^+$ and $\dot{\Theta}^+$. This means that the Lagrange brackets among the physical variables, which  form a matrix, can be computed on the gauge slice. Furthermore, we find that this 
matrix is invertible, indicating that the choice of the gauge $\Theta^+=0$ is 
 a proper one, and this inverse gives  the Dirac brackets  for the physical variables we want. 

To make the logic clear, we shall demonstrate this in  slightly abstract notations and 
then apply the formulas to our specific  system to give concrete results. 
In the following, we collectively denote the bosonic and the fermionic variables by $X^\abar$ and $\Theta^\albar$ respectively and write our Lagrangian as 
\begin{align}
L &= {1\over 2e} J^\abar J^\abar \period
\end{align}
Here  $J^\abar$ represent  the components of the appropriate currents  along
 the bosonic coset space, {\it not yet gauge-fixed}. $J^\abar$ is utmost linear in 
$\dot{X}$ or $\dot{\Theta}$. 
The momenta conjugate to $X^\abar$ 
and $\Theta^\albar$ are given by 
\begin{align}
P_\abar &= {\del L \over \del \dot{X}^\abar} = {1\over e} 
{\del J^\bbar \over \del \dot{X}^\abar} J^\bbar \comma \label{Pbose} \\
P_\albar &= {\del L \over \del \dot{\Theta}^\albar} = {1\over e} 
{\del J^\bbar \over \del \dot{\Theta}^\albar} J^\bbar \period 
\label{Pfermi}
\end{align}
As is true for our system, we consider the case where the matrix $M_\abar{}^\bbar \equiv \del  J^\bbar/\del \dot{X}^\abar $ is invertible. Then from (\ref{Pbose}) 
 we can solve for $J_\bbar$ as
\begin{align}
J_\bbar &=e (M^{-1})_\bbar{}^\abar  P_\abar \period \label{JintP}
\end{align}
Putting this into  (\ref{Pfermi}) we obtain  $d_\albar=0$ 
where 
\begin{align}
d_\albar & = P_\albar - {\del J^\bbar \over \del \dot{\Theta}^\albar} 
(M^{-1})_\bbar{}^\abar  P_\abar  \period \label{fconst}
\end{align}
As $d_\albar$'s  consist  of basic phase space variables only,  they represent 
 fermionic constraints. 
(There are also bosonic constraints generated by the presence of the einbein, 
 but as they are not  important  in the ongoing  analysis, we will discuss them later.)
Now we make an assumption that (\ref{fconst}) are the only fermionic 
 constraints and that  the $\kappa$-gauge symmetry generated by 
  half of them can be fixed by setting $\Theta^+_I=0$, where $\Theta^+_I$ represents   an appropriate  half of $\Theta^\albar$. The remaining ``physical" part will be denoted by $\Theta^-_I$. This assumption is quite reasonable since the degrees of freedom of the system should not differ from the flat case.  In any case, it will be supported  by the results of our anlysis. 

We now wish to compute the Lagrange bracket among  the physical phase space variables  $(X^\abar, P_\abar, \Theta^-_I)$.  Let us  first give a brief review of this 
 bracket for the case without constraints.  Let $( p_i, q^i)$ 
be  a basis  of the $2N$ dimensional phase space, including 
 fermionic variables.  As we have used left derivative to 
 define the fermionic momenta, the appropriate definition of the Poisson bracket
for aribtrary  functions $F$ and $G$ is 
\begin{align}
\{ F , G \}_P = (-1)^{|i|}\frac{\partial F}{\partial_R q^i}
\frac{\partial G}{\partial_L {p_i}}
- \frac{\partial F}{\partial_R {p_i}}\frac{\partial G}{\partial_L q^i} \period
\end{align}
Here  the subscripts $L$ and $R$ refer to  the left and the right derivatives
 resectively and $|i|=0\, (1)$ for the bosonic (fermionic) variable. 
Let $\{z_\mu\} {\scriptstyle \mu=1\sim 2N }$ be functions of $(p,q)$ which 
form a complete basis of  the phase space. Then, the Lagrange bracket 
 between $z_\mu$ and $z_\nu$  is given by\footnote{It should be clear that 
the   subscript $L$ on the bracket stands for ``Lagrange" and not for ``Left".}
 \begin{align}
( z_{\mu} , z_{\nu})_L = 
(-1)^{|i|}\frac{\partial p^i}{\partial_L z_{\nu}}
\frac{\partial q_i}{\partial_R z_{\mu}} 
- \frac{\partial q_i}{\partial_L z_{\nu}}
\frac{\partial p^i}{\partial_R z_{\mu}}  \period \label{Lagbra}
\end{align} 
One can easily show that the Lagrange bracket is the inverse of the Poisson bracket 
 in the sense 
\begin{align}
(z_{\mu} , z_{\nu})_L \{ z_{\mu} , z_{\rho}\}_P = \delta_{\nu\rho} 
\period \label{LagPoi}
\end{align} 

Next consider the case with constraints.  As  said before  we assume that 
by adding the  gauge-fixing costraints  $\Theta^+_I=0$, the total set of 
$M$  constraints $(d_\albar, \Theta^+_I)$ can be made second class. 
  It can then be shown that  if we take $(d_\albar, \Theta^+)$ themselves to 
 be among the $z_\mu$ functions, the counter part of the relation (\ref{LagPoi}) 
holds  for the $2N-M$ physical variables in the form
 \begin{align}
 \sum^{2N - M}_{\mubar = 1} (z_{\mubar} , z_{\nubar})_L \{ z_{\mubar}, z_{\rhobar}\}_D = \delta_{\nubar\rhobar},
 \qquad  (\nubar,\rhobar = 1, \cdots 2N - M)\comma 
 \end{align}
where $\Dcom{z_\mubar}{z_\rhobar}$  is the Dirac bracket. This means that 
  the Dirac bracket for the physical variables can be computed as the inverse 
 of the their Lagrange bracket. 

Let us compute the Lagrange brackets $(z_{\mubar} , z_{\nubar})_L $ 
more explicitly by taking 
 \begin{align}
 (p_i, q^i) &= (X^\abar, P_\abar, \Theta^-_I, \Theta^+_I,  P_\albar) \comma 
\label{piqi}\\
 (z_{\mu}) &= (z_\mubar,   \Theta^+_I, d_\albar) \comma 
\qquad z_\mubar = (X^\abar, P_\abar, \Theta^-_I) \comma \label{zmu}
 \end{align}
Note that $(p,q )$ and $(z_\mu) $ differ only by $P_\albar \leftrightarrow d_\albar$
and  $P_\albar$ can be regarded as a function of $z_\mu$ 
by the use of the relation (\ref{fconst}), namley $P_\albar 
= d_\albar + (\del J^\bbar/\del \dot{\Theta}^\albar)M^{-1}_\bbar{}^\abar 
P_\abar$.  Although the general definition of  the Lagrange bracket 
 is already given in (\ref{Lagbra}),  let us display it again for the physical 
 variables $z_\mubar$, as it will be very important:
\begin{align}
(z_{\mubar} , z_{\nubar})_L  &=
(-1)^{|i|}\frac{\partial p^i}{\partial_L z_{\nubar}}
\frac{\partial q_i}{\partial_R z_{\mubar}} 
- \frac{\partial q_i}{\partial_L z_{\nubar}}
\frac{\partial p^i}{\partial_R z_{\mubar}}  \period \label{physLagbra}
\end{align}
We now make two simple but crucial observations about this formula.
 First, since  $\Theta^+$
 is not among the variables $z_\mubar$  the derivative 
 with respect to $\Theta^+$ cannot appear on the right hand side. 
Second, its conjugate $P_{\Theta^+}$ can only appear with  $\Theta^+$ 
in the form like  $(\del P_{\Theta^+}/\del z_\mubar) ( \del \Theta^+/\del z_\nubar)$. But this vanishes because  $ \del \Theta^+/\del z_\nubar=0$. Hence 
 $P_{\Theta^+} = \del L/\del \dot{\Theta}^+$ never appears in 
(\ref{physLagbra}).  Combining, we find that the Lagrange bracket for the physical variables can be computed without knowing the dependence on  $\Theta^+$ and  $\dot{\Theta}^+$. 
In other words, the knowledge of the relevant quantities 
{\it on the gauge slice} is sufficient to compute it.  
It is straightforward to evaluate the right hand side of (\ref{physLagbra}) 
for the choice (\ref{piqi}) and (\ref{zmu}) and 
 obtain the following useful formulas:
\begin{align}
(X^\abar, X^\bbar)_L &=  (P_\abar, P_\bbar)_L =0 \comma \qquad 
(X^\abar, P_\bbar)_L = \delta^\abar_\bbar \comma \label{Lbone}\\
(X^\abar, \Theta^-_I)_L &= 
-\left({\del P_{\Theta^-_I} \over  \del X^\abar}\right)_{\Theta^+=0} 
\comma \qquad (P_\abar, \Theta^-_I)_L = 
-\left({\del P_{\Theta^-_I} \over  \del P_\abar}\right)_{\Theta^+=0} 
\comma \label{Lbtwo}\\
(\Theta^-_I, \Theta^-_J)_L &=
  -\left({\del P_{\Theta^-_I} \over \del_L (\Theta^-_J)}
+{\del P_{\Theta^-_J} \over \del_R (\Theta^-_I)} \right) _{\Theta^+=0} 
\period \label{Lbthree}
\end{align}

Having explained our  method of computation, let us apply it to the 
superparticle case at hand. From the explicit form of the MC 1-forms 
given in (\ref{MCP}) $\sim$ (\ref{MCD}),  it is straightforward to compute the fermionic momenta on the gauge slice, namely $P_\albar 
=\left. (\del J^\bbar/\del \dot{\Theta}^\albar)(M^{-1})_\bbar{}^\abar 
P_\abar\right|_{\Theta^+=0}$. 
The result is 
 \begin{align}
 P_{\theta^{i}} &= \frac{i}{2}\theta^-_iP_- \comma \qquad 
 P_{\theta_i} = \frac{ i}{2}\theta^{i}P_- \comma \\
 P_{\eta^{i}} &= \frac{ i}{2}\eta_ie^{2\phi}P_- \comma \qquad 
 P_{\eta_i} = \frac{ i}{2}\eta^{i}e^{2\phi}P_- \period
 \end{align}   
Substituting them into the general formulas (\ref{Lbone}) $\sim$ (\ref{Lbthree}), 
we readily obtain the explicit form of the Lagrange brackets. In the formulas 
below, we use the notation 
 $x^{\underline{a}}$ to represent
 the set of  bosonic coordinates $(x^+, x^-, x, \xbar, \phi (=x^4), y^\Ap)$. Then the results can be written as
 \begin{align}
(x^{\underline{a}} , x^{\underline{b}} )_L &= (P_{\underline{a}},P_{\underline{b}})_L = 0 \comma \qquad 
(x^{\underline{a}} , P_{\underline{b}} )_L = \delta^{\underline{a}}_{\underline{b}} \\
 (x^{\underline{a}} , \theta^{i} )_L & =
 (x^{\underline{a}} , \theta_i )_L =  0\comma  \\
 (x^{\underline{a}} , \eta^{i})_L  &= -i \eta_ie^{2\phi}P_-\delta^{\underline{a}}_4 \comma \qquad 
  (x^{\underline{a}} , \eta_i)_L  =  -i \eta^{i}e^{2\phi}P_-\delta^{\underline{a}}_4 \comma \\
 (P_{\underline{a}} , \theta^{i})_L &=  -\frac{i }{2}\theta_i
\delta^-_{\underline{a}}  \comma \qquad 
 (P_{\underline{a}} , \theta_i)_L = -\frac{i }{2}\theta^{i}
\delta_{\underline{a}} ^- \comma \\
 (P_{\underline{a}} , \eta^{i})_L &
 = -\frac{i }{2}\eta_ie^{2\phi}
\delta_{\underline{a}} ^- \comma \qquad 
 (P_{\underline{a}} , \eta_i)_L 
 = -\frac{i }{2}\eta^{i}e^{2\phi}
\delta_{\underline{a}} ^-\comma  \\
 (\theta^{i} , \theta_j)_L &
 = -i P_-\delta^i_j\comma  \qquad 
 (\eta^{i} , \eta_j)_L 
 = -i e^{2\phi}P_-\delta^i_j \period 
 \end{align}  
From these expressions we confirm that the Lagrange bracket  $(z_\mubar, z_\nubar)_L$ as a matrix is  invertible, justifying our assumption made earlier. 
Actual inversion is quite easy and we obtain  the Dirac brackets as 
 \begin{align}
 \{ x^{\underline{a}} , P_{\underline{b}} \}_D &= \delta^{\underline{a}}_{\underline{b}}\comma  \\
 \{ x^{\underline{a}} , \theta^{i} \}_D &=  -\frac{1}{2P_-}\theta^{i}
\delta^{\underline{a}}_-\comma  \qquad 
 \{ x^{\underline{a}} , \theta_i \}_D =  -\frac{1}{2P_-}\theta_i
\delta^{\underline{a}}_-\comma  \\
 \{ x^{\underline{a}} , \eta^{i} \}_D &=  -\frac{1}{2P_-}\eta^{i}
\delta^{\underline{a}}_-\comma  \qquad 
 \{ x^{\underline{a}} , \eta_i \}_D =  -\frac{1}{2P_-}\eta_i
\delta^{\underline{a}}_-\comma  \\
 \{ P_{\underline{a}} , \eta^{i} \}_D &= \eta^{i}
\delta_{\underline{a}}^{4}\comma \qquad 
 \{ P_{\underline{a}} , \eta^-_i \}_D = \eta_i
\delta_{\underline{a}}^{4}\comma  \\
 \{ \theta^{i} , \theta_j \}_D &= \frac{i }{P_-}\delta^i_j\comma  \qquad 
 \{ \eta^{i} , \eta_j \}_D = \frac{i }{P_-}e^{-2\phi}\delta^i_j 
\period 
 \end{align}
It is evident that, just as in the flat case, the variables  $\theta$ and  $\eta$  no longer satisfy the canonical bracket relations. The experience for the flat case suggests
that we should form  the following combinations:
   \begin{align}
   S_{\theta^{i}} & =  \sqrt{P_-}\theta^{i}, \qquad 
   S_{\theta_i} =  \sqrt{P_-}\theta_i,\qquad 
   S_{\eta^{i}}  =  \sqrt{P_-}e^{\phi}\eta^{i}, \qquad 
   S_{\eta_i} =  \sqrt{P_-}e^{\phi}\eta_i \period
   \end{align}
The extra factor  $e^\phi$ is introduced  for the $\eta$'s to make their conformal weight  equal to that of  $\theta$'s. Then  it is not difficult to check that these 
new variables commute with the bosonic variables and satisfy the canonical 
 bracket relations 
\begin{align}
\Dcom{S_{\theta^{i}}}{ S_{\theta_j} } &=\Dcom{S_{\eta^{i}}}{ S_{\eta_j} }= i \delta^i_j \comma \qquad \mbox{Rest$=0$} \period 
\end{align}

Finally, let us  discuss the remaining first class bosonic constraints which so far have 
 been suppressed. They  are  
 $P_e=0, T=0$, where $P_e$ is the momentum conjugate to the 
einbein $e(\tau)$ and $T$ is the reparametrization generator. 
$T$ is related to the Hamiltonian $H$ by $eT=H$. The calclulation of the canonical 
Hamiltonian in the semi-light-cone gauge is straightforward but slightly
 cumbersome. Since all the terms in the action (\ref{genaction}) are quadratic in the time derivative, we have 
\begin{align}
H &= L = {1\over 2e} \left( 2J_B^+J_B^-+ 2 J_B^x J_B^\xbar + (J_B^\phi)^2
 + J_B^\Ap J_B^\Ap \right)\period 
\end{align}
 What is non-trivial is the step of expressing 
 the relevant MC 1-forms in terms of the phase space variables.  This  requires the 
 explicit evaluation of the formula  (\ref{JintP}), with the aid of some formulas of  the Appendix A.  After some computations we get
   \begin{equation}
H = e\left( 2e^{2\phi}(P_+P_- + P_xP_{\bar{x}}) + 
\frac{1}{2}( P_{\phi}^2 + \lhat^2 
+ (S_{\eta}^2)^2) - 2S_{{\eta}_i}l^i{}_jS_{{\eta}^j}\right)
\period 
\end{equation}
Here $l^i{}_j$ and $\lhat^2$ are, respectively, the orbital part of the $su(4)$
 genrator and the associated  quadratic Casimir operator, which are
  discussed  in the Appendix A. 
As our primary goal in this paper is to compute the physical spectrum of the system, 
we will  fix the gauge symmetries generated by these constraints as well by imposing  the   conditions
\begin{equation}
\xi(\tau) = e(\tau) - 1\comma  \qquad 
\chi(\tau) = x^+(\tau) - \tau \period \label{lccond}
\end{equation} 
In this bonafide light-cone gauge, all the constraints become second class. 
The addition of the conditions above requires us to modify the Dirac bracket
slightly. However, it is easy to see that only the brackets with $P_+$ need to be 
changed  and its effect can be implemented within the unmodified  Dirac bracket 
by replacing   $P_+$ by the expression 
\begin{align}
P_+ &=- {P_x P_\xbar \over P_-} - {e^{-2\phi}  \over 4P_-}
\left(P_\phi^2 + (\lhat^2) + (S_\eta^2)^2 \right) + e^{-2\phi}S_{{\eta}_i}l^i{}_jS_{{\eta}^j} \comma 
\end{align}
which is obtained by solving the constraint $T=0$ explicitly. With this understanding 
 we need not modify our Dirac bracket. Due to the gauge condition (\ref{lccond})
 $x^+ (=\tau)$ becomes non-dynamical and the $\tau$-evolution of any 
 function $F$  is generated by 
 the light-cone Hamiltonian $H^{l.c.} = -P_+$ as $dF/d\tau = \del F/ d\tau
 + \Dcom{F}{H^{l.c.}}$. 
 \section{Quantization and the quantum  Noether charges  }
\subsection{Quantization}
Now that we have clarified the phase space formulation of the dynamics of a superparticle in $AdS_5 \times S^5$ and obtained the Dirac brackets, it is  straightforward to quantize our  system: We simply   replace $i \Dcom{\ }{\ }$ by 
 the equal time quantum commutator $\com{\ }{\ }$.  In addition, for convenience 
 we will introduce simplified notations for the quantized fermionic variables. 
The new variables $S^i, S_i, \Stil^i, \Stil_i$ are defined as 
\begin{align}
S^i &= i S_{{\theta}^i} \comma \qquad S_i = i S_{{\theta}_i}
\comma \qquad 
\Stil^i = i S_{{\eta}^i} \comma \qquad \Stil_i = iS_{{\eta}_i}  
\period 
\end{align}
Then the commutation relations of the fundamental variables take the form
\begin{align}
\com{x}{P_x} &= \com{\xbar}{P_\xbar} =\com{x^-}{P_-} =\com{\phi}{P_\phi} = i  \comma \quad \com{y^{A'}}{P_{B'}} = i \delta^{A'}_{B'} \comma \\
\acom{S^i}{S_j} &= \acom{\Stil^i}{\Stil_j} = \delta^i_j  \comma 
\qquad \mbox{Rest $=0$} \period
\end{align}
\subsection{Derivation of the Noether charges and their quantization }
As explained at the beginning of section 2.3, our strategy for the solution of the 
quantum  dynamics of a  superparticle is to make use of the 
 realization of the $psu(2,2|4)$ symmetry of the system. In preparation for this goal, 
we shall derive  in this subsection  the Noether charges for this symmetry and 
quantize them in a systematic manner. 

To begin, let us first recall how we can find these charges by the Noether method 
in the configuration space. Before gauge fixing, the $PSU(2,2|4)$ transformation
 acts on the supercoset representative $G$ in the manner
\begin{align}
G \longrightarrow f_0 G h \comma 
\end{align}
where $f_0= e^{\ep_0}$
 is an element of $PSU(2,2|4)$ and   $h$ is a compensating 
$SO(4,1) \times SO(5)$ transformation which depends  on $X, \Theta$ and $f_0$. 
For a global  $\ep_0$ the action is invariant under this 
 transformation.  To derive the Noether charges, one makes the parameter local 
 and replaces  $f_0$ by  $ f = e^{\ep(\tau)}$.  Then, the bosonic coset
 part of  the MC 1 form $J_B$ gets transformed as
\begin{align}
J_B \longrightarrow & \left[ (fGh)^{-1} d(fGh) \right]_B = 
\left[ h^{-1} G^{-1} (f^{-1} df) Gh + h^{-1} G^{-1} dG h + h^{-1}dh
 \right]_B 
\nn\\
&= [G^{-1} \dot{\ep}^{\hat{A}} T_{\hat{A}} G]_B + \cdots \comma 
\end{align}
where $T_{\hat{A}}$  denotes  the generator  of $psu(2,2|4)$ and 
 the ellipses stand  for terms independent of $\dot{\ep}$ or higher order in $\ep$. 
From this one can obtain the Noether charge $\mbQ_{\Ahat}$ 
corresponding to $T_{\hat{A}}$ as\footnote{
 We will denote the Noether charges in   bold blackboard 
style, to distinguish them from the generic $psu(2,2|4)$ generators. }
\begin{align}
\mathbb{Q}_\Ahat & = \frac{1}{e\dot{\epsilon}}\Str (\delta_{\Ahat} J)_B
J_B = \frac{1}{e}(G^{-1}T_\Ahat G)^{\underline{a}}J^{\underline{a}}_B
\notag \\
&  = \frac{1}{e}\left(  \frac{1}{\sqrt{2}}(-(\delta_\Ahat L_K^{a}) + (\delta_\Ahat L_P^{a}))J_B^{a}
 +(\delta_\Ahat L_B^D)J_B^4 + \frac{i }{2}(\delta_\Ahat L_B)^j{}_i(\gamma^{A'})^i{}_jJ_B^{A'} \right). \label{Noeformula}
\end{align}
In this formula,   $\delta_\Ahat L$'s are the  expressions which appear in 
 the expansion
\begin{equation}
G^{-1}T_\Ahat G = (\delta_\Ahat L_K^{a})K^a + (\delta_\Ahat L_P^a)P^a 
+ \cdots .
\end{equation}
Now when one fixes the gauge, the naive  transformation law
 is no longer valid. Since 
 the $PSU(2,2|4)$ transformations in general do  not preserve the gauge, 
one must perform  appropriate compensating gauge transformations 
in order to keep the gauge condition intact.  For a superstring in the 
$AdS_5\times S^5$  background, the  cumbersome  task of finding such transformations was accomplished in \cite{Nishimura:2006ad}. 

Next  let us discuss the case of the phase space formulation. 
Compared to the procedure in the configuration space just reviewed, the computation  in the phase space formulation is much simpler. In particular, we need not 
 find the compensating transformations explicitly once we have the proper Dirac bracket. This is because the Dirac bracket, by definition, automatically provides 
 the requisite  projection onto the gauge slice. Moreover, it solves another related problem at the same time. 
This is the apparent  problem of  ambiguities  one encounters 
 when one tries to convert  the configuration space expressions  into those 
 in the phase space. Namely, any combination of the constraints can be added 
 in the conversion formula. It should however  be clear that as far as the computations using the Dirac brackets are concerned this is of no problem. Under the Dirac bracket, the constranits can be set strongly to zero and the result is unambiguous. 
 Therefore, the formula (\ref{Noeformula}) 
 can be used as it is, with the replacement $J_{B \una} = e ( ({\del J /\del \dot{X}})^{-1})_\una{}^\unb P_\unb$. Hence, we have the formula 
\begin{align}
\mbQ_\Ahat &= (G^{-1}T_\Ahat G)^\una  
( ({\del J /\del \dot{X}})^{-1})_\una{}^\unb P_\unb \comma 
\end{align}
which can be evaluated directly on the gauge slice. After involved but straightforward 
calculations, we obtain all the classical Noether charges and check that 
 they satisfy the $psu(2,2|4)$ algebra under the Dirac bracket. 
As we shortly  display  the quantum version of the Noether charges in full detail, 
 the list of the classical charges so obtained is  relegated to the  Appendix B to 
 avoid redundancy\footnote{
In \cite{Metsaev:2000yu} classical Noether charges 
 for the superstring case were obtained ( although  some of them were not displayed explicitly).  When restricted to the superparticle mode,  they agree with our results. However, 
in contrast to our direct systematic derivation, they  resorted to some indirect reasoning for obtaining the dynamical generators. 
}. 

The remaining problem is to find the quantum representation of the charges. 
The main task is to fix the ordering of the operators and it  can be done by requiring 
 the  realization of the 
  hermiticity properties (\ref{hermite}) and the  closure of the $psu(2,2|4)$ algebra. 
One simplifying fact  is that the $\tau$-dependence is generated by 
 a  member of the algebra, namely by $H^{l.c.} = -P_+$. Thus, we can 
work at the  time slice  $\tau=0$ and later recover  the $\tau$-dependence. 

The first step is to impose the Hermiticity conditions (\ref{hermite}) on  
  the Noether charges.  The rules  of conjugation for the basic  variables are
\begin{equation}
(x^{\underline{a}})^{\dagger} = x^{\underline{a}}, \qquad 
(P_{\underline{a}})^{\dagger} = P_{\underline{a}},\qquad 
(S^{i})^{\dagger} = S_i, \qquad  (\tilde{S}^i)^{\dagger} = \tilde{S}_i
\comma 
\end{equation}
where, as before, $x^{\underline{a}}$ represent  all the bosonic coordinates. 
We find that this  process  fixes the operator  orderings  for  $\mathbb{P}^+, \mathbb{P}^x, \mathbb{P}^{\bar{x}}, \mathbb{Q}^{-i}, \mathbb{Q}^-{}_i,
\mathbb{K}^+, \mathbb{D}, \mathbb{J}^{+-}, \mathbb{J}^{x\bar{x}}, \mathbb{J}^{+x}, 
\mathbb{J}^{+\bar{x}}$ and $\mathbb{J}^i{}_j$. 
Next, demand that $\{ \mathbb{Q}^{-i}, \mathbb{Q}^-{}_j\} $ becomes 
 proportional to $\delta^i{}_j$.  This turns out to fix the ordering of  $\mathbb{Q}^-$.  In this calclulation, we made  use of  the following relation satisfied by  the 
orbital part of the quantized $su(4)$ generators $l^i{}_j$:
\begin{equation}
l^i{}_jl^j{}_k = \frac{1}{4}\lhat^2\delta^i{}_k + 2l^i{}_k 
\comma \qquad \lhat^2 \equiv  l^i{}_jl^j{}_i \period \label{llid}
\end{equation}
This identity, which appeared in \cite{Metsaev:1999gz}, is discussed in 
 the Appendix A and will be of importance again in the next section. 
The third step is to demand that $\{ \mathbb{Q}^+, \mathbb{S}^- \} $ satisfy the correct algebra.  This condition fixes  the ordering  of $\mathbb{S}^-$ because the 
 ordering ambiguity  is proportional to the  operator of the form 
$S/\sqrt{P_-}$.
Up to this point  all the fermionic generators have been fixed. 
The ordering of the remaining bosonic operators are then determined by  requiring 
the proper closure of the $psu(2,2|4)$ algebra.  It turned out that
 only a part of the algebra  was needed to fix the ambiguities but we have verified 
the remaining 
 part as well for a good consistency check. 
Finally the forms of the operators  at general $\tau$ can be computed  by 
 the unitary transformation $\calO(\tau) = e^{\tau \mbP^-} \calO(0) 
 e^{-\tau \mbP^-}$, which actually terminates  at order $\tau^2$. 

We now display all the quantum Noether charges thus obtained, 
regarded  as the generators of  the four dimensional 
 superconformal algebra.   For convenience we  use  the 
following notations:
\begin{align}
z &\equiv e^\phi \comma \qquad 
 N_S = S^iS_i \comma  \qquad \NStil=\Stil^i\Stil_i \comma 
\qquad  S\cdot \Stil = S^i \Stil_i  \period 
\end{align}
$z$ is the coordinate along  the direction normal to the boundary of $AdS$, 
$N_S$ and $\NStil$ are  the number operators for $S^i$ 
and  $\Stil^i$ respectively. 

First, the translation generators are given by\footnote{
Although we use the variable 
$z$ instead of $\phi$, it should be remembered that the hermiticity is still defined 
with respect to the $\phi$ variable.  In particular, the hermitian 
momentum is $P_\phi =-i \del_\phi = -i(1/z)\del_z$. }
\begin{align}
\mbP^x &= iP_\xbar \comma  \qquad 
\mbP^\xbar = iP_x   \comma \qquad 
\mbP^+ = iP_-  \comma \\
\mbP^- & 
= {i \over 4P_-} \Bigl[ -4 P_x P_\xbar + \del_z^2 - {1\over z} \del_z 
+ {1\over z^2} ( -3-\lhat^2 + 4 \NStil -\NStil^2 + 4 l^m{}_k \Stil^k \Stil_m ) 
\Bigr] \period
\end{align}
The special conformal generators are 
\begin{align}
\mbK^x &= -i z^2 P_\xbar 
 + x \left( z \del_z + ix^- P_- + ix P_x  + \half (\NS -\NStil + 3)
\right)   + iz S\cdot \Stil -\tau \mbJ^{-x} \comma  \\
\mbK^\xbar &= -i z^2 P_x 
 + \xbar \left( z \del_z + ix^- P_- + i\xbar P_\xbar  + \half (-\NS +\NStil + 3)
\right) 
 + iz \Stil\cdot S -\tau \mbJ^{-\xbar} \comma \\
\mbK^+ &= {1\over i} (z^2 +x\xbar) P_-  
 +\tau (  z\del_z + ixP_x + i\xbar P_\xbar +1+  \tau
 \mbP^-)   \comma \\
\mbK^- &= (x\xbar -z^2) \mbP^- + \xbar \mbJ^{-x} + x \mbJ^{-\xbar}
+x^- z \del_z +i  (x^-)^2 P_- +2 x^-  \nn\\
&\qquad  + {i\over 4P_-} \Bigl[ -2z \del_z -1 + 2\NStil -\NStil^2 -2\NS + \NS^2 
 +4(\Stil\cdot S)( S \cdot \Stil) \nn\\
&\qquad + 4 l^k{}_m 
( \Stil^m \Stil_k -S^m S_k) 
-4z ( P_\xbar \Stil\cdot S + P_x S \cdot \Stil) \Bigr]  \period
\end{align}
The dilatation operator takes the form 
\begin{align}
\mbD &= -z \del _z - (i x^-P_-  + ix P_x  + i\xbar P_\xbar ) - {3 \over 2} -\tau 
\mbP^-  \period
\end{align}
As for the Lorentz generators  we get
\begin{align}
\mbJ^{+-} &= -i x^- P_- + \half + \tau \mbP^- \comma \\
\mbJ^{x\xbar} &= -i\xbar P_\xbar + ix P_x + \half (\NS -\NStil) \comma \\
\mbJ^{+x} &= -ix P_- + i\tau P_\xbar \comma \\
\mbJ^{+\xbar} &= -i\xbar P_- + i\tau P_x \comma \\
\mbJ^{-x} &= -x \mbP^- + ix^- P_\xbar -{P_\xbar \over 2P_-}
( \NS + \NStil -1) +{i\over \sqrt{P_-}} S^k \mbQ^-_k \comma \\
\mbJ^{-\xbar} &= -\xbar \mbP^- + ix^- P_x + {P_x \over 2P_-} 
( \NS + \NStil +1) +{i\over \sqrt{P_-}} \mbQ^{-k} S_k \period
\end{align}
The $su(4)$ generators  $\mbJ^i{}_j$ consist of the orbital part $l^i{}_j$ 
and the spin part $M^i{}_j$:
\begin{align}
\mbJ^i{}_j &= l^i{}_j + M^i{}_j \period
\end{align}
$l^i{}_j$ and $M^i{}_j$  separately satisfy the same $su(4)$ algebra as $\mbJ^i{}_j$. 
The explicit form of $l^i{}_j$  is rather involved and is 
discussed in the Appendix A,  together with its properties. On the other hand, 
  the spin part is quite simple and is given by 
\begin{align}
 M^i{}_j & = S^i S_j -{1\over 4} \delta^i_j \NS +\Stil^i \Stil_j -{1\over 4} \delta^i_j \NStil   \period \label{spingen}
\end{align}
Finally the supertranslation and the superconformal generators are given by 
\begin{align}
\mbQ^{+i} &= i \sqrt{P_-} S^i  \comma \\
\mbQ^+_i &= -i \sqrt{P_-} S_i  \comma \\
\mbQ^{-i} &= {i \over 2\sqrt{P_-}} \Bigl[ 2 P_x S^i  -\del_z \Stil^i 
 + {1\over z} \left(  \Stil^i  (\NStil -1)  -2 l^i{}_k \Stil^k \right)   \Bigr] \comma \\
\mbQ^-_i &= {-i \over 2\sqrt{P_-}} \Bigl[ 
2 P_\xbar S_i + \del_z \Stil_i + {1 \over z} \left( \Stil_i (\NStil -3) 
 -2 \Stil_k l^k{}_i \right) \Bigr]  \comma \\
\mbS^{+i} &= -i \sqrt{P_-} \left( z \Stil^i + i\xbar S^i \right)
 +i \tau \mbQ^{-i} \comma \\
\mbS^+_i &= i \sqrt{P_-} \left( z \Stil_i -ix S_i \right) -i\tau \mbQ^-_i  \comma \\
\mbS^{-i} &= {-i \over 2 \sqrt{P_-}} \Bigl[ 2z P_\xbar \Stil^i 
 -2\Stil^i (S \cdot \Stil) - S^i (z \del_z + \NS+1)+ 2 l^i{}_k S^k \Bigr] 
+ix^- \mbQ^{+i} + ix \mbQ^{-i}  \comma \\
\mbS^-_i &= {i \over 2 \sqrt{P_-}} \Bigl[ 2z P_x \Stil^i 
 -2\Stil^i (\Stil \cdot S) +S_i (z \del_z - \NS +5) + 2 l^k{}_i S_k \Bigr] 
-ix^- \mbQ^+_i - i\xbar \mbQ^-_i  \period
\end{align}
Here we should remark  that, although not all the generators were explicitly displayed, in \cite{Metsaev:1999gz} 
Metsaev ingeniously wrote down essentially the same form 
of  generators,   without  systematic derivations. 

Having derived the complete set of quantum generators  for the 
$psu(2,2|4)$ superconformal algebra,  we are now ready to 
 study the physical states which form  unitary irreducible representations of 
 this  concrete system. 
\section{Solution of the  superconformal primary states}
Our strategy for studying  the spectrum and  other quantum properties of the 
superparticle in $AdS_5\times S^5$ 
 is to make maximal use of the representation theory of  its  symmetry algebra. 
The general  theory of the representations of $psu(2,2|4)$ algebra has been 
fairly well-developed\cite{Dobrev:1985qv, Dobrev:1985qz,Gunaydin:1984fk,Minwalla:1997ka,Dolan:2002zh} and the classification of all the unitary irreducible representations are known.  They include special  short and semi-short ``BPS" representations, which have been realized in various parts of AdS/CFT correspondence.  With such  knowledge at hand, the problem 
 we wish to solve is  to find precisely 
  which  representations can be  realized in the Hilbert space where  the generators 
 of $psu(2,2|4)$ are realized in the specific form given  in the previous section. 
In this section we will give a complete answer  to this problem by  constructing all possible 
 superconformal primary states, including  their explicit  wavefunctions. 
\subsection{Dilatation (D) scheme  and energy (E) scheme }
To begin, it is important to discuss the two commonly used 
schemes of describing the representations 
 of the conformal group $SO(4,2)$\footnote{These two schemes were
extensively discussed in \cite{Dolan:2002zh}.}. 
 They will be called the E-scheme and 
 the D-scheme and are based on the following maximal subgroup decomposition:
\begin{align}
\mbox{$E$-scheme} \qquad & SO(4,2) \supset SO(2)_{E}
 \times SU(2)_L\times SU(2)_R \comma \\
\mbox{$D$-scheme} \qquad & SO(4,2) \supset SO(1,1)_{D} 
\times SL(2,C) \times \overline{SL(2,C)} \period
\end{align}
Recall that in our convention, the embedding coordinates are labeled as $(X^{-1}, X^0, X^1,X^ 2, \allowbreak X^3,X^4)$ with the signature $(-,-,+,+,+,+)$. In the E-scheme,  $SO(2)_E$ acts  on the coordinates $(X^{-1},X^0)$, while $SU(2)_L \times SU(2)_R \simeq SO(4)$ rotates $(X^1,X^2,X^3,X^4)$. The generator of $SO(2)_E$ is 
 the hermitian AdS energy, to be denoted by $\mbE$.  In terms of the generators in the light-cone basis, it is given by
\begin{align}
\mbE &={1\over i} T^{0,-1} =  {1\over 2i} (\mbP^+ -\mbP^--\mbK^+ +\mbK^-) \period \label{defE}
\end{align}
  On the other hand, in the D-scheme 
$SO(1,1)_D$ acts on $(X^{-1},X^4)$ and the Lorentz group $ SL(2,C) \times \overline{SL(2,C)}\simeq SO(3,1)$ acts on $(X^0,X^1,X^2,X^3)$. 
The generator of $SO(1,1)$ is the dilatation operator $\mbD$. In our convention
 it is anti-hermitian. 

Clearly, these two schemes are related by the exchange  $X^0 \leftrightarrow X^4$, which is generated by the anti-hermitian boost operator
\begin{align}
R \equiv T^{40} = \half (\mbP^+ - \mbP^- +\mbK^+ -\mbK^-) \period 
\label{defR}
\end{align}
In fact, by using the basic commutation relations of $SO(4,2)$, it is easy to see that 
$\mbD$ is mapped to  $\mbE$ by a  similarity transformation of the form 
\begin{align}
& V \mbD  V^{-1}  = \mbE  \comma \qquad V = \e^{i(\pi/2) R}  \period \label{simtransV}
\end{align}
Of course one can map any generator $\calO$  of $psu(2,2|4)$ by this similarity 
transformation and we denote it by 
\begin{align}
V \calO V^{-1} = \hat{\calO} \period \label{simtransV2}
\end{align}
In this 
notation, $\mbE = \mbDhat$. 
As it is a similarity transformation, this mapping preserves  the structure
 of the superconformal algebra.  
However,  it is important to note that  it is a {\it non-unitary} transformation and hence it does not  preserve the norm.  Therefore, to obtain a unitary (hence normalizable ) representation, one must choose an appropriate scheme. 

As we expect to be able to reproduce the supergravity result, the proper scheme should  be the E-scheme, with {\it real values for  the AdS energy $E$}. This will be confirmed 
 in the subsequent sections.  Therefore, as for the $SO(4,2)$ part, we will label 
the states by the eigenvalue of $\mbE$ and those of the  Cartan generators
  $\mbJ^3_{L,R}$ of the $su(2)_{L,R}$ algebras. In terms of the light-cone basis generators, 
$\mbJ^3_{L,R}$ are given by 
\begin{align}
\mbJ^3_L &= \half (\mbH_1 + \mbH_2) \comma \qquad 
\mbJ^3_R = \half (\mbH_1 - \mbH_2) \comma 
\end{align}
where
\begin{align}
\mbH_1 &= \mbJ^{x\xbar} =\mbJhat^{x\xbar} \comma \qquad 
\mbH_2 = {i\over 2} ( \mbP^++\mbP^-+\mbK^++\mbK^-) 
= -\mbJhat^{+-} \period
\end{align}
The eigenvalues  of the hermitian operators $\mbH_1$ and $\mbH_2$ will be denoted by $h_1$ and $h_2$ respectively. 

The non-unitary nature of the similarity transformation above manifests itself most 
conspicuously   in the following fact. Suppose $\ket{E}$ is  a unit-normalized energy eigenstate with real non-zero  eigenvalue $E$, {\it i.e.} $\mbE \ket{E} = E\ket{E}$. Then, $\mbD 
(V^{-1} \ket{E}) = V^{-1} (V \mbD V^{-1}) \ket{E} =
V^{-1} \mbE \ket{E} =  E(V^{-1} \ket{E})$. 
In other words,  the state $V^{-1} \ket{E}$ is an eigenstate of an anti-hermitian
generator $\mbD$ with real eigenvalue  $E$. As is well-known, this can only happen 
if  $V^{-1} \ket{E}$  is of zero-norm.  Indeed the norm of this state, which is 
$\bra{E} V^{-2} \ket{E}$,  vanishes since 
as one  can easily show that $V^{-2} \mbE  = -\mbE V^{-2}$.  
In the context of AdS/CFT, this phenomenon is consistent with the fact 
 that the gauge-invariant 
composite operators in the super-Yang-Mills theory carry real eigenvalues 
with respect to the anti-hermitian dilatation operator\footnote{We thank R. Janik for a discussion on this point.}.  However, it is rather non-trivial that, group theoretically,  what corresponds  to a physical  CFT operator with  a definite dilatation charge is  a zero-norm state   on the AdS side, which is hard to interpret 
physically. It would be interesting to clarify this structure more deeply. 
\subsection{Superconformal primary conditions }
We now formulate the problem of finding the superconformal primary states  in the E-scheme  more explicitly.  In this scheme, 
the superconformal primary state $\Psiket$ is 
characterized  by the following 16 conditions 
\begin{align}
\mbShat^{\pm i} \Psiket = 0 \comma \qquad \mbShat^\pm_i \Psiket =0 
\comma 
\end{align}
where $\mbShat^{\pm i} = V \mbS^{\pm i} V^{-1}$ and $\mbShat^\pm_i 
 = V \mbS^\pm_i V^{-1}$. 
From the form of $R$ given in (\ref{defR}) and the basic commutation relations 
listed in section 2, we obtain
\begin{align}
\com{R}{\mbS^{\pm i}} &= \mp {i \over 2} \mbQ^{\mp i} \comma 
\qquad \com{R}{\mbS^\pm_i } = \pm  {i \over 2} \mbQ^\mp_i \comma \\
\com{R}{\mbQ^{\pm i}} &= \mp {i \over 2} \mbS^{\mp i} \comma 
\qquad \com{R}{\mbQ^\pm_i } = \pm  {i \over 2} \mbS^\mp_i  \period
\end{align}
By repeatedly applying these commutation relations, we can easily 
obtain a formula such as $e^{i\theta R} \mbS^{+i} e^{-i\theta R}
 = \mbS^{+i} \cos(\theta/2) + \mbQ^{-i} \sin(\theta/2)$, etc. Setting 
$\theta =\pi/2$, we get the superconformal generators in the E-scheme  as 
\begin{align}
\mbShat^{+i} &= \ovsqtwo (\mbS^{+i} + \mbQ^{-i} ) \comma \qquad 
\mbShat^{-i} = \ovsqtwo (\mbS^{-i} - \mbQ^{+i} ) \comma \\
\mbShat^+_i  &= \ovsqtwo (\mbS^+_i - \mbQ^-_i ) \comma \qquad 
\mbShat^-_i  = \ovsqtwo (\mbS^-_i + \mbQ^+_i )  \period 
\end{align}
\nullify{
\begin{align}
\mbKhat^x &= \half (\mbK^x -\mbP^x  -i(\mbJ^{+x} -\mbJ^{-x})) \nn\\
\mbKhat^\xbar &= \half ( \mbK^\xbar-\mbP^\xbar -i(\mbJ^{+\xbar} -\mbJ^{-\xbar})) \nn\\
\mbKhat^+&= \half (\mbK^+-\mbP^- -i(\mbJ^{+-} - \mbD)) \nn\\
\mbKhat^-&= \half (\mbK^- -\mbP^+-i(\mbJ^{+-} + \mbD)) \nn
\end{align}
}
The desendants of the irreducible  representation are generated from the 
superconformal primary state by 
the repeated action of the E-scheme version of the supertranslation generators $\mbQhat$'s, which can be obtained in an entirely similar manner. They are given by 
\begin{align}
\mbQhat^{+i} &= \ovsqtwo (\mbQ^{+i} + \mbS^{-i} ) \comma \qquad 
\mbQhat^{-i} = \ovsqtwo (\mbQ^{-i} - \mbS^{+i} ) \comma \\
\mbQhat^+_i  &= \ovsqtwo (\mbQ^+_i - \mbS^-_i ) \comma \qquad 
\mbQhat^-_i  = \ovsqtwo (\mbQ^-_i + \mbS^+_i )  \period
\end{align}
The $SO(4,2)$ quantum numbers $(E, h_1, h_2)$ carried by these $\mbQhat$ operators are
\begin{align}
&\mbQhat^{+i}:\ \left( \half, \half, -\half\right) \comma \qquad \mbQhat^{-i}:\
\left(\half, -\half, \half\right) \comma \\
&\mbQhat^+_i:\ \left( \half, -\half, -\half\right) \comma \qquad \mbQhat^-_i:\
 \left( \half, \half, \half \right)
\end{align}

Because of the relation $\acom{\mbShat}{\mbShat} \sim \mbKhat$, 
the conformal primary conditions $\mbKhat^a \Psiket =0$ are automatically 
 satisfied by the superconformal primaries. In this sense, we need not impose them separately. However, as they will be useful in the subsequent analysis, we will briefly discuss their explicit forms. By applying  the similarity transformation (\ref{simtransV2})
to $\mbK^a$, we easily obtain the desired counterparts  in the E-scheme:
\begin{align}
\mbKhat^x &= \half (\mbK^x -\mbP^x  -i(\mbJ^{+x} -\mbJ^{-x})) \comma \\
\mbKhat^\xbar &= \half ( \mbK^\xbar-\mbP^\xbar -i(\mbJ^{+\xbar} -\mbJ^{-\xbar})) \comma \\
\mbKhat^+&= \half (\mbK^+-\mbP^- -i(\mbJ^{+-} - \mbD)) \comma \\
\mbKhat^-&= \half (\mbK^- -\mbP^+-i(\mbJ^{+-} + \mbD)) \period 
\end{align}
On (super)conformal primaries these generators vanish.  It will be convenient to 
 express this by  the notation\footnote{This useful notation 
was introduced in \cite{Metsaev:2002vr}.} 
$\mbKhat^a \approx 0$. Then, from $\mbKhat^\pm \approx 0$
 we can express $\mbP^-$ and $\mbK^-$, which are among the most 
 complicated generators,  in terms of other simpler generators. Explicitly, 
\begin{align}
\mbP^- &\approx \mbK^+-i (\mbJ^{+-} -\mbD ) \comma \\
\mbK^- &\approx \mbP^+ + i (\mbJ^{+-} +\mbD ) \period
\end{align}
If we apply these relations to the AdS energy operator given in (\ref{defE}), 
we obtain a relation
\begin{align}
\mbE  &\approx -i (\mbP^+-\mbK^++ i\mbJ^{+-}) \period
\label{EPKJ}
\end{align}
This will be of use  in the next subsection. 
\subsection{Allowed unitary  highest weight representation for the {\bfall $su(4)$}
 sector}
When expressed in terms of the basic quantum variables of the superparticle, 
the superconformal primary conditions in the E-scheme 
formulated in the previous subsection  are  actually quite involved even at $\tau=0$
and cannot be analyzed as they stand. 

We now make two observations which will simplify the situation. 
The first observation is that the dependence
 on the $S^5$ coordinates and the derivatives with respect to them is 
 only through the generators $l^i{}_j$  of the orbital part of $su(4)$. 
This means that the Casimir operator $\lhat^2$ 
commutes with  all the generators of the $psu(2,2|4)$ and hence we can analyze 
the possible representations of this orbital part independently. 

The second observation is that the aforementioned quadratic relation
 (\ref{llid}) satisfied by $l^i{}_j$ is quite useful  for our analysis. 
For the present purpose, we display it again in the following form:
\begin{align}
\calL^i{}_j &\equiv 
l^i{}_k l^k{}_j  - {1\over 4} \lhat^2\delta^i_j  - 2l^i{}_j  =0 \period 
\label{llrel}
\end{align}
Existence of such  product relations among the generators dictate that the structure
 of the representation module is correspondingly restricted.  

To illustrate this in the simplest possible setting, 
 consider the $su(2)$ algebra realized by the generators $(J^3, J^\pm )$ 
made out of two sets of 
 fermionic oscillators $(b_i, b^i)_{i=1,2}$ with the 
anticommutation relations  $\acom{b_i}{b^j} = \delta^i_j, \acom{b^i}{b^j} = \acom{b_i}{b_j} =0$ in the following way:
\begin{align}
J^3&= \half (b^1 b_1 -b^2 b_2)  \comma \qquad J^+ = b^1 b_2 \comma 
\quad J^-= b^2 b_1 \period
\end{align}
As usual the  highest weight module is constructed by acting $J^-$ to a highest 
weight state $\ket{j}$ satisfying $J^+\ket{j} =0$ and $J^3 \ket{j} = j\ket{j}$. 
But because of the special form of the generators,  there is an obvious 
 product relation $J^-J^-=0$. This clearly restricts the allowed representations 
 to be  utmost two dimensional. Indeed for this system  two 
singlet representations with the highest weight states $\ket{0}$ and $b^1b^2\ket{0}$ and one doublet representation with the highest weight state $b^1\ket{0}$ are the only possible representations. 

Let us  now examine the consequences of the product relation (\ref{llrel}). The fastest way  is to use the explicit realization of $l^i{}_j$ in terms of the Chevalley basis 
 generators $(H_\ihat, E^\pm_\ihat)_{\ihat=1,2,3}$, which satisfy the 
commutation relations 
\begin{align}
\com{H_\ihat}{H_\jhat} &= 0 \comma \qquad 
\com{E^+_\ihat}{E^-_\jhat} = \delta_{\ihat\jhat} H_\jhat  \\
\com{H_\ihat}{E^\pm_\jhat} &= \pm K_{\jhat \ihat} E^\pm_\jhat \comma 
\qquad K_{\jhat \ihat}=\mbox{Cartan matrix} \period 
\end{align}
It reads 
{\small 
\begin{align}
& l^i{}_j \nn\\
&=
\mativ{{1\over 4}(3H_1+2H_2+H_3) & E^+_1 & \com{E^+_1}{E^+_2} & 
\com{E^+_1}{\com{E^+_2}{E^+_3} } }{E^-_1 & {1\over 4}(-H_1+2H_2+H_3) & E^+_2 & \com{E^+_2}{E^+_3}}{-\com{E^-_1}{E^-_2}
& E^-_2 &-{1\over 4}(H_1+2H_2-H_3)& E^+_3  }{
\com{E^-_1}{\com{E^-_2}{E^-_3} }& -\com{E^-_2}{E^-_3}&
E^-_3 &-{1\over 4}(H_1+2H_2+3H_3) }  \label{Chevalley}
\end{align}
}
Since we are interested in what highest weight representations 
 are allowed, we apply 
(\ref{llrel}) onto a highest weight state 
$\ket{\lam_1, \lam_2, \lam_3}$, where $\lam_i (\ge 0)$
 denote  the Dynkin weights.  Such a state is characterized by 
$E^+_i \ket{\lam_1, \lam_2, \lam_3} =0$ and 
$H_i \ket{\lam_1, \lam_2, \lam_3} = \lam_i \ket{\lam_1, \lam_2, \lam_3}$. 

Consider first the equation  $\calL^2{}_1 \ket{\lam_1, \lam_2, \lam_3} =0$. 
Using the Chevalley basis expressions, the left hand side can be easily computed and the resultant equation is equivalent to 
\begin{align}
(\lam_1 + 2\lam_2 + \lam_3 + 2) E^-_1  \ket{\lam_1, \lam_2, \lam_3} =0 \period
\end{align}
Since the coefficient in front is non-vanihsing, we get $E^-_1  \ket{\lam_1, \lam_2, \lam_3} =0$.  It means that the state must be a singlet of $su(2)$ along 
 the direction 1 and hence we must have $\lam_1=0$. In an entirely similar manner, 
the relation 
$\calL^4{}_3 \ket{\lam_1, \lam_2, \lam_3} =0 $ dictates $\lam_3=0$. One can then check that the rest of the relations are automatically satisfied and 
do not lead to any further restrictions. Summarizing, the allowed highest 
weight states  for the orbital part of $su(4)$ are of the form 
\begin{align}
\ket{0,l,0} \comma \qquad l=0,1,2, \ldots \period
\end{align}
From (\ref{Chevalley}) we can easily work out the action of the generators 
$l^i{}_j$ on this state.  In particular, the following relations will be useful later:
\begin{align}
l^i{}_j \lket &=0\quad  \mbox{for $i < j$} \comma \qquad l^2{}_1\lket = l^4{}_3 \lket =0 \comma  \nn\\
l^1{}_1 \lket &= l^2{}_2 \lket = -l^3{}_3 \lket = -l^4{}_4 \lket = \half l 
\lket \comma \label{lonlket} \\
\lhat^2 \lket &= l(l+4) \lket \period  \nn
\end{align}
It should be noted that the analysis above can be regarded as  an efficient algebraic 
means  for performing the  harmonic analysis  on $S^5$. 

Next, we wish to apply a  similar anaysis of the representation of $su(4)$  to   the entire Hilbert space, including the spin part.  Here we encounter a difficulty: The 
total generators  $\mbJ^i{}_j = l^i{}_j + M^i{}_j$ do not satisfy any  simple 
product relation such as (\ref{llrel}).  The reason for this is simple. The spin part of 
 the Hilbert space is generated by the action of the eight creation operators $S^i, \Stil^i$  and hence consists of  $2^8$ states. These states 
 fall    into a large number of different highest weight representations. Therefore, tensored  with the orbital part, 
the representations in the entire Hilbert space 
 are not so severely restricted as  in the  orbital case. 

Fortunately, however, product relations similar to (\ref{llrel}) do exist {\it on 
superconformal primary states}. Specifically, consider the following linear combination of superconformal primary conditions: 
\begin{align}
\sqrt{P_-} \left(z \Stil_j \mbShat^{+i} - z \Stil^i \mbShat^+_j  -S_j \mbShat^{-i} + S^i \mbShat^-_j \right) \approx 0 \period
\end{align}
Substituting the explicit form of the $\mbShat$ generators  and 
making use of the formula (\ref{EPKJ}), we obtain, 
after some computation, a useful relation 
\begin{align}
&\Stil^i \Stil_j (1-\NStil) + S^i S_j (1-\NS) - \Stil^i (S \cdot \Stil) S_j - S^i (\Stil \cdot S) \Stil_j  \nn\\
& + \half \delta^i_j (\NS + \NStil) + l^i{}_k(S^k S_j + \Stil^k \Stil_j) 
 + l^k{}_j(S^i S_k +\Stil^i \Stil_k) -2l^i{}_j -\mbE \delta^i_j \approx 0 \period
\end{align}
This in turn can be used to compute $\mbJhat^i{}_k \mbJhat^k_j$. 
After some computation the result can be put into the form
\begin{align}
\calJ^i_j &\equiv 
\mbJhat^i{}_k \mbJhat^k{}_j - {1\over 4} \mbJhat^2 \delta^i_j 
- \left( 4-{N \over 2} \right) \mbJhat^i{}_j \  \approx \  0 \comma \label{JJrel}
\end{align}
where 
\begin{align}
&  N \equiv \NS + \NStil\comma   \qquad (N=0,1, \ldots , 8) \comma \\
&  \mbJhat^2 \equiv  \mbJ^i{}_k \mbJ^k{}_i
\ \approx \ 4\mbE + \lhat^2 -{1\over 4} (N-4)^2 + 4 \period 
\label{Jhatsq}
\end{align}
Since the relation (\ref{JJrel}) has the form quite similar  to  (\ref{llrel}), 
we can repeat the  analysis for each value of $N$. In fact as each $\mbShat$ 
 operator has  a definite $N$-number,  it is not possible to form 
a superconformal primary state 
by making a linear combination of   non-superconformal primary states 
 with   different $N$-numbers. This means that the basic irreducible 
 superconformal primary state must carry a definite value of  $N$. 

In this analysis, in order to narrow down the allowed representations, 
it is useful to impose the unitarity requirement at the same time. 
From the hermiticity property of $\mbQ$'s and $\mbS$'s, we easily find the 
hermiticity of their counterparts in the E-scheme, namley 
$\mbQhat$'s and $\mbShat$'s,  as follows:
\begin{align}
(\mbQhat^{+i})^\dagger &= \mbShat^-_i \comma \quad 
(\mbQhat^+_i)^\dagger = -\mbShat^{-i} \\
(\mbQhat^{-i})^\dagger &= -\mbShat^+_i 
\comma \quad (\mbQhat^-_i)^\dagger = \mbShat^{+i} \period
\end{align}
Now let $\Psiket$ be a superconformal primary state, which is annihiliated by 
$\mbShat$'s. Then for a unitary representation, we get a so-called 
unitarity bound  by 
\begin{align}
\langle \Psi | \{ \mbShat^-_i, \mbQhat^{+i} \} \Psiket
 = | \mbQhat^{+i}\Psiket |^2 \ge 0 \comma \label{unitaritybd} 
\end{align}
and further  bounds using  other pairs. One can easily evaluate the anticommutator
such as $\{ \mbShat^-_i, \mbQhat^{+i} \}$ by transforming the known 
result for $\{ \mbS^-_i, \mbQ^{+i} \}$. If the $su(4)$ part of $\Psiket$ 
 is taken to be $\ket{\lam_1, \lam_2, \lam_3}$, we can evaluate the left-hand-side
 of (\ref{unitaritybd}) explicitly.  In this way, one can obtain 
a useful bound such as $E \ge \lam_1 + \lam_2 + \lam_3$, which can be used 
 to eliminate a number of possible representations during the analysis 
 of (\ref{JJrel}). Since the detail is somewhat involved  we relegate it 
 to the Appendix C. 

The outcome  of this anaysis is that the allowed highest 
weight states for the $su(4)$ sector 
can only be  of the following three types:
\begin{align}
&(i)\quad \Omlket = S^1 \Stil^1 S^2 \Stil^2 \ket{0} \otimes \ket{0,l,0}
\comma \quad l=0,1,2, \ldots \comma \label{Omlket}\\
&(ii)\quad \ket{{\rm vac}} = \ket{0} \otimes \ket{0,0,0} \comma \label{vac} \\
&(iii)\quad \ket{{\rm fvac}} =
S^1\Stil^1 S^2 \Stil^2 S^3 \Stil^3 S^4 \Stil^4 \ket{0} \otimes  \ket{0,0,0}
\period \label{fvac}
\end{align}
 The first factor of the tensor product 
is the spin part and the second factor is the orbital part. 
The symbols ``vac" and ``fvac"  signify the vacuum and the filled-vacuum nature of 
 the spin part.  In the next subsection, we will show that only the 
states of type $(i)$ will lead to  the proper 
normalizable superconformal primary states. 
\subsection{Solutions and properties  of the superconformal primary states }
\subsubsection{Solutions  of the superconformal primaries at {\bfall $\tau=0$}}
We are now ready to solve the superconformal primary conditions explicitly. 
In this subsection, we will concentrate on  the solutions at $\tau=0$. 

First consider the states built upon $\Omlket$ given in  (\ref{Omlket}). 
We will write a state of this type as
\begin{align}
\Psilket = \Phi_l (z, P_-, P_x , P_\xbar) \Omlket \period 
\end{align}
On  such  a state,  the supercharge operators  simplify 
substantially. Since half of the fermionic oscillators are excited in $\Omlket$, 
below  we will split the $su(4)$ index $i$ as $i=(\al, \alhat)$, where 
$\al=1,2$ and $\alhat =3,4$. Then,  the fermionic oscillators act
on $\Omlket$ as
\begin{align}
S_\alhat \Omlket &= \Stil_\alhat \Omlket =S^\al\Omlket = \Stil^\al \Omlket
=0 \comma \\
\NS \Omlket &= \NStil \Omlket = 2 \Omlket \comma  \\
\Stil \cdot S \Omlket &= S \cdot \Stil \Omlket=0 \period
\end{align}
As for the structures involving  $l^i{}_k$, using (\ref{lonlket}) we get
\begin{align}
l^\alhat {}_k S^k \Omlket &= -{l\over 2} S^\alhat \Omlket  \comma \qquad 
l^\al {}_k S^k \Omlket =0 \comma \\
 l^\alhat{}_k \Stil^k  \Omlket &=-{l\over 2} \Stil^\alhat \Omlket
\comma \qquad  l^\al{}_k \Stil^k  \Omlket  =0 \comma 
\\
S_k l^k{}_\al  \Omlket &={l\over 2} S_\al \Omlket 
\comma \qquad S_k l^k{}_\alhat  \Omlket =0 \comma  \\
\Stil_k l^k{}_\al   \Omlket &= {l\over 2} \Stil_\al \Omlket \comma \qquad 
\Stil_k l^k{}_\alhat   \Omlket =0 \period  
\end{align}
Applying these results to  the supercharges 
$\mbQ$'s and $\mbS$'s,  we find that they  simplify considerably and 
effectively reduce to  the following forms on $\Psilket$:
\begin{align}
\mbQ^{\pm \al } &= 0 \comma \qquad 
\mbQ^{\pm}_\alhat = 0 \comma  \label{vanishQ}\\
\mbQ^{+\alhat } &= 
 i \sqrt{P_-} S^\alhat \comma \qquad 
\mbQ^+_\al  = -i \sqrt{P_-} \Stil_\al \comma \label{Qplusalhat} \\
\mbQ^{-\alhat } &= {i \over 2\sqrt{P_-}} \left( 2 P_x S^\alhat  -\left( \del_z 
-{l+1\over z} \right)\Stil^\alhat 
   \right) \comma \label{Qminusalhat} \\
\mbQ^-_\al  &= {-i \over 2\sqrt{P_-}} \left(
2 P_\xbar S_\al  + \left(\del_z - {l+1 \over z} \right)\Stil_\al \right) \comma 
\end{align}
\begin{align}
\mbS^{\pm \al} &=0 \comma \qquad 
\mbS^\pm_\alhat =0 \comma \label{vanishS}\\
\mbS^{+\alhat} &= -i \sqrt{P_-} \left( z \Stil^\alhat -{\del \over \del P_\xbar}
 S^\alhat \right) \comma \label{Splusalhat}
  \\
\mbS^+_\al &= i \sqrt{P_-} \left( z \Stil_\al +{\del \over \del P_x}  S_\al \right) \comma  \\
\mbS^{-\alhat} &= {-i \over 2 \sqrt{P_-}} \Bigl[ 2z P_\xbar \Stil^\alhat  
  -(z \del_z +l+ 3) S^\alhat   \Bigr] 
-{\del \over \del P_-} \mbQ^{+\alhat} -{\del \over \del P_x}  \mbQ^{-\alhat} \comma \label{Sminusalhat}\\
\mbS^-_\al  &= {i \over 2 \sqrt{P_-}} \Bigl[ 2z P_x \Stil_\al
  +(z \del_z +l+3) S_\al  \Bigr] 
+{\del \over \del P_-} \mbQ^+_\al +{\del \over \del P_\xbar}  \mbQ^-_\al   \period 
\end{align}
Note that,  combining (\ref{vanishQ}) and (\ref{vanishS}),  
the following half of the superconformal primary conditions in the E-scheme 
are automatically satisfied:
\begin{align}
\mbShat^{\pm \al} \Psilket=0\comma \qquad \mbShat^\pm_\alhat \Psilket =0 \period
\end{align}
Similarly, we see that the half of the supercharges in the E-scheme annihiliate 
$\Psilket$:
\begin{align}
\mbQhat^{\pm \al} \Psilket=0 \comma \qquad \mbQhat^\pm_\alhat \Psilket =0
\period 
\end{align}
This means that {\it all the highest weight representations of this type are half BPS}. 

We now impose the remaining superconformal primary conditions one by one
to determine the form of $\Phi_l$ . 

First, consider the condition 
\begin{align}
0 &= \sqtwo \mbShat^{+\alhat} \Psilket = (\mbS^{+\alhat} + \mbQ^{-\alhat})
\Psilket  \nn\\
&= {i \over 2\sqrt{P_-}} \Biggl[ 2\left( P_x + P_- {\del \over \del P_\xbar}
\right)S^\alhat   -\left( \del_z -{l+1 \over z} + 2P_-z \right)\Stil^\alhat 
\Biggr] \Psilket \period 
\end{align}
From the coefficient of $S^\alhat$ and $\Stil^\alhat$, we get two first order 
differential equations:
\begin{align}
& \left( P_x + P_- {\del \over \del P_\xbar}
\right) \Phi_l =0  \comma \\
& \left( \del_z -{l+1 \over z} + 2P_-z \right)\Phi_l =0 \period
\end{align}
The first equation  determines the $P_\xbar$ dependence and gives
\begin{align}
\Phi_l &= f_1(z, P_x, P_-) \exp\left( -{P_x P_\xbar \over P_-} \right) \period 
\end{align}
The second equation  on the other hand determines the dependence on $z$ and gives%
\begin{align}
\Phi_l &= f_2(P_-, P_x, P_\xbar) \exp \left( -z^2 P_- \right) z^{l+1} \period
\end{align}
Combining, we get
\begin{align}
\Phi_l &= f(P_-) \psi  \comma  \\
 \psi &= \exp\left(  -{P_x P_\xbar \over P_-}  -z^2 P_- \right) z^{l+1} \period
\label{psieq}
\end{align}

 Next consider the following condition
\begin{align}
0 &= \sqtwo \mbShat^{-\alhat} \Psilket = ( \mbS^{-\alhat} - \mbQ^{+\alhat})
\Psilket  \period 
\end{align}
After some calculations one obtains 
\begin{align}
( \mbS^{-\alhat} - \mbQ^{+\alhat})
\Psilket  
&= -{i \over \sqrt{P_-}}
\Biggl[ (1+z^2) P_-  -\left( l+\half\right) 
 -{P_x P_\xbar \over P_-} 
+P_- {\del \over \del P_-} \Biggr]S^\alhat \Psilket  =0 \period
\end{align}
Plugging in the form of $\Phi = f(P_-) \psi$ above,  we get the equation for $f(P_-)$
of the form 
\begin{align}
{\del \over \del P_-} f &= \left( {l+\half \over P_-} -1\right) f \period 
\end{align}
This is readily solved to give
\begin{align}
f &= C_l e^{-P_-} P_-^{l+(1/2)}  \comma 
\end{align}
where $C_l$ is a constant. 

Finally, one can easily check that the remaining conditions $0 = \sqtwo \mbShat^+_\al  \Psilket = ( \mbS^+_\al - \mbQ^-_\al)\Psilket $ and $0 = \sqtwo \mbShat^-_\al
  \Psilket = ( \mbS^-_\al + \mbQ^+_\al)
\Psilket $ are satisfied automatically. 

Summarizing, we have found that  upon $\Omlket$ 
a unique superconformal primary state exists  for each $l$,
 which takes the form\footnote{For  some special  states belonging to {\unboldmath $ l=0$} multiplet, 
Metsaev obtained the bosonic part of the  wave function in \cite{Metsaev:2002vr}.}
\begin{align}
\Psilket &=  C_l  \exp\left(  -{P_x P_\xbar \over P_-}  -(z^2+1) P_- \right) z^{l+1}
P_-^{l+(1/2)} \nn\\
& \quad \times  S^1\Stil^1 S^2 \Stil^2 \ket{0}\otimes \ket{0,l,0}  
\comma \qquad l=0,1,2, \ldots \period \label{Psilketsol}
\end{align}
The quantum numbers of this state are read off  by acting $\mbE, \mbJ^3_{L,R}$ and  $\mbJhat^2$ on this state. We obtain
\begin{align}
\mbE \Psilket &= E_l \Psilket \comma \qquad E_l =l+2  \comma \\
\mbJ^3_{L,R} \Psilket &= 0 \comma \\
\mbJhat^2 \Psilket &= (l+2)(l+6) \Psilket \period
\end{align}

The descendants  belonging to the highest weight representation 
  are produced  by operating the 
$8$  supercharges  $\mbQhat^{\pm 3,4}, 
 \mbQhat^\pm_{1,2}$ (and the momentum operators $\mbPhat$'s) on  
$\Psilket$. Mapping  the basic commutation relations  $\com{\mbD}{\mbQ} = \half \mbQ$ to the E-scheme, we obtain $\com{\mbE}{\mbQhat} = \half \mbQhat$.  So each time we act by a  $\mbQhat$,   the AdS energy is raised by $\half$ unit. 
For example, one of the first excited states is of the form  $\mbQhat^{+3} \Psilket = \sqrt{P_-/2} S^3 \Psilket $, which carries the energy $l+(3/2)$. From the form of $\Psilket$, the 
dimension of the representation  (up to the action of $\mbPhat$'s) is readily seen to be given by 
\begin{align}
2^8 \times \dim[0,l,0] &= {64 \over 3} (l+1)(l+2)^2 (l+3) \period
\end{align}
It is precisely that of the $\half$ BPS superconformal multiplets  of 1-particle states  realized in type IIB supregravity\cite{Kim:1985ez,Gunaydin:1984fk}.
 In the AdS/CFT context, these states  correspond to the single trace operator   $\trace (\phi^{\{I_1} \phi^{I_2}   \cdots \phi^{I_{l+2}\}} )$ and its descendants in $N=4$ Super-Yang-Mills theory. 

 In supergravity, one often speaks of the ``mass" formula, which expresses the eigenvalue of the  D'Alembertian for the $AdS$ space. Although it is not a genuine  invariant of the entire $psu(2,2|4)$ algebra, we expect it to be related  to  the value of the quadratic Casimir operator $\half T^{AB} T_{AB}$ of  the $so(4,2)$ subalgebra. Indeed, on $\Psilket$ we find the well-known formula 
\begin{align}
\half T^{AB} T_{AB} \Psilket &= E_l(E_l-4) \Psilket  \period
\end{align}

The states $\Psilket$ are normalizable in the standard quantum mechanical sense\footnote{Note that we are dealing with  a quantum wave function of a particle and 
not with a supergravity field. So the measure should be taken to be the one 
 appropriate for  quantum mechanical interpretation.} 
namely with respect to the integration measure which respects the hermiticity 
 of the basic variables.  Explicitly, the  squared norm of $\Psilket$  is given by 
\begin{align}
\int_0^\infty { dz\over z} 
  \int_0^\infty dP_- \int_{-\infty}^\infty 
dP_1 \int_{-\infty}^\infty  dP_2\,   \bra{\Psi_l}\Psi_l\rangle &= |C_l|^2 
 {(l+1) (l!)^2 \over 2^{2l+4} }\pi \period
\end{align}
Let us make two remarks.  First in our scheme the measure for $z$ variable should be taken to be   $dz/z =d\phi$, since $\phi$ and its conjugate $P_\phi$ were regarded as   the basic hermitian variables.  Second, the range of $P_-$ should be taken to be the 
 semi-infinte interval $\left[ 0, \infty\right]$. The reason is that $\Psilket$ vanishes 
at both ends of this interval and this insures the hermiticity of $x^-$ and $P_-$.

Having constructed the series of states $\Psilket$ upon $\Omlket$, 
 let us now consider the  superconformal primary state built upon $\ket{{\rm vac}}$  given in (\ref{vac}). On $\ket{{\rm vac}}$, 
$S_i$, $\Stil_i$ and all the orbital $su(4)$  generators $l^i{}_j$ vanish.  This immediately leads to $\mbQ^+_i = \mbQ^-_i = \mbS^+_i = \mbS^-_i =0$. The remaining  supercharges effectively take the form
\begin{align}
\mbQ^{+i} &= i\sqrt{P_-} S^i \comma \\
\mbQ^{-i} &={i \over 2\sqrt{P_-}} \left( 2P_x S^i -\left(\del_z  +{1\over z}\right)  \Stil^i \right)  \comma \\
\mbS^{+i} &= -i\sqrt{P_-} \left(z \Stil^i  -{\del \over \del P_\xbar}S^i \right)
\comma  \\
\mbS^{-i} &=  {-i \over 2 \sqrt{P_-}} \Bigl[ 2z P_\xbar \Stil^i
  -(z \del_z +1) S^i   \Bigr] 
-{\del \over \del P_-} \mbQ^{+i} -{\del \over \del P_x}  \mbQ^{-i} \period
\end{align}
Note that the form of $\mbQ^{+i}$ and $\mbS^{+i}$ are the same as 
 (\ref{Qplusalhat}) and (\ref{Splusalhat}) for $i=\alhat$, while  $\mbQ^{-i}$ and $\mbS^{-i}$ coincide with (\ref{Qminusalhat}) and (\ref{Sminusalhat}) for $i=\alhat$  if we  set $l=-2$. With  this in mind,  going through the analysis praralell to the previous case,  we easily find  that the primary state on $\ket{{\rm vac}}$ is of the form
\begin{align}
\ket{\Psi_{{\rm vac}}} &= \Phi_{l=-2} \ket{{\rm vac}} 
 \propto  \exp\left(  -{P_x P_\xbar \over P_-}  -(z^2+1) P_- \right)
z^{-1} P_-^{-3/2} \ket{{\rm vac}} \period \label{Psivac}
\end{align}
As indicated, the part other than $\ket{{\rm vac}}$ is identical to 
$\Psilket$  for  $l=-2$ and hence  this state has $E=0$.
 It is however  no longer normalizable: 
The integral over $z$ for $\bra{\Psi_{{\rm vac}} } \Psi_{{\rm vac}} 
\rangle $ behaves like  $\sim \int  dz/z^3$ near $z=0$ and is divergent. 

The analysis for the state built upon $\ket{{\rm fvac}}$ given in 
(\ref{fvac}) is very  similar. The wave function is identical to (\ref{Psivac}) above 
 except $\ket{{\rm vac}}$ is replaced by $\ket{{\rm fvac}}$. Such a  state 
is also not normalizable. 
\subsubsection{Complete solution  at arbitrary {\bfall $\tau$} }
Although the spectrum and the quantum numbers can be read off  from $\Psilket$ 
at $\tau=0$, it is of interest to compute the full wave
 function $\ket{\Psi_l(\tau)}$ at   arbitrary $\tau$ in order to see the 
profile of the wave function in the $AdS$ space and gain physical understanding.
 As already explained  previously, it is obtained 
from the solution $\Psilket$  at $\tau=0$ by the unitary transformation 
\begin{align}
\ket{\Psi_l(\tau)} &= e^{\tau \mbP^-} \Psilket\period
\end{align}
Upon $\Psilket$ the operator $\mbP^-$ simplifies to 
\begin{align}
\mbP^-&= {i \over 4P_-} \left(D_z ^{(l)} 
-4P_x P_\xbar \right) \comma \nn\\
D_z^{(l)} &\equiv  \del_z^2 -{1\over z} \del_z -{l^2-1\over z^2} \period
\end{align}
Nevertheless, the direct evaluation of the unitary transformation 
 above is still difficult.  A standard trick is to convert it to a Schr\"odinger equation 
by differentiating it with respect to $\tau$. We obtain 
$\del_\tau \ket{\Psi_l(\tau)} = \mbP^- \ket{\Psi_l(\tau) }$, which 
 can be rewritten as 
\begin{align}
 4\left( {1\over i}P_- \del_\tau + P_x P_\xbar\right)\ket{\Psi_l(\tau)}
 = D_z ^{(l)}  \ket{\Psi_l(\tau)} \period \label{Schrodeq}
\end{align}
Now one can easily check that the eigenfunction of the differential 
 operator $D_z ^{(l)} $ is given by 
\begin{align}
 D_z ^{(l)} f_l(\be, z) &= -\be^2  f_l(\be, z) \comma \\
 f_l(\be, z) &= z J_l(\be z)  \comma 
\end{align}
where $J_l(x)$ is the standard Bessel function of order $l$.  Moreover, 
 a very useful integation formula involving $J_l(\be z)$ exists. It reads
\begin{align}
\int_0^\infty d\be e^{-\be^2/4P_-} \be^{l+1} J_l(\be z)
 &= z^l 2^{l+1} P_-^{l+1} e^{-z^2 P_-}  \period \label{Jint}
\end{align}
It allows us to express the solution  $\Psilket$ given in (\ref{Psilketsol}) as 
the following  integral
\begin{align}
\Psilket &= \int_0^\infty d\be \psi_0(\be) f_l(\be, z)  \Omlket 
\comma \\
 \psi_0(\be) &=C_l 2^{-(l+1)} P_-^{-1/2}   \be^{l+1}  \exp\left(-{\be^2\over 4P_-}\right) 
\exp\left( -{P_x P_\xbar \over P_-} -P_-\right) \period 
\end{align}
This suggests that we should seek the solution of the 
Schr\"odinger equation in the form 
\begin{align}
\ket{\Psi_l(\tau)} &=  \int_0^\infty d\be \psi(\be,\tau) f_l(\be, z)  \Omlket 
\comma  \label{Psitau}
\end{align}
where the function $\psi(\be,\tau)$ should satisfy the initial condition 
$\psi(\be,\tau=0) =\psi_0(\be)$. Putting (\ref{Psitau}) into (\ref{Schrodeq}), 
one obtains a simple first order differential equation with respect to $\tau$ 
for  $\psi(\be,\tau)$, which can be readily  solved. The solution 
 obeying the initial condition is 
\begin{align}
\psi(\be, \tau) &= \psi_0(\be)  \chi(\be, \tau) \comma \\
\chi(\be,\tau) &= \exp\left( {-i\tau \be^2\over 4P_-}\right)
 \exp\left( -i {P_x P_\xbar \over P_-} \tau \right)  \period
\end{align}
Since $\chi(\be, \tau)$  is a Gaussian in $\be$, we can perform 
 the integral (\ref{Psitau})  by using the formula (\ref{Jint}) again. 
In this way, we obtain $\ket{\Psi_l(\tau)}$ in  a closed form.  Reinstating the 
  original relation  $\tau = x^+$, it reads 
\begin{align}
\ket{\Psi_l(z,  P_x, P_\xbar, P_-, x^+)}
&= C_l \left({z \over 1+ix^+} \right)^{l+1} P_-^{l+(1/2)} \nn\\
& \quad \times \exp \left( -{P_x P_\xbar \over P_-} (1+ix^+) -P_- 
 -{z^2 P_- \over 1+ix^+} \right)  \Omlket \period
\end{align}

To get a feel  for  this wave function, let us perform the Fourier transform  to go to  the full coordinate representation. The transforms  with respect to $P_x$ and $P_\xbar$ are standard and yield 
\begin{align}
\ket{\Psi_l(z, x, \xbar,  P_-, x^+)}
&= C_l  z^{l+1} (1+ix^+)^{-(l+2)} P_-^{l+(3/2)}
 e^{- \al P_-}\Omlket \comma \\
\al &\equiv {x\xbar + z^2 +1 + ix^+ \over 1+ix^+} \period
\end{align}
As for the transform with respect to $P_-$, we employ the general formulas for 
the semi-infinite interval, namely 
\begin{align}
\tilde{f}(x) &= \int_0^\infty  {dp \over \sqrt{2\pi}} e^{i px} f(p) \comma \qquad 
f(p) = \int_{-\infty}^\infty {dx \over \sqrt{2\pi}} e^{-ipx} \tilde{f}(x) \period
\end{align}
In this way  the full coordinate representation is obtained as 
\begin{align}
\ket{\Psi_l(z, x, \xbar,  x^-, x^+)}
&= C_l  z^{l+1} (1+ix^+)^{-(l+2)}   \int_0^\infty {dP_-\over \sqrt{2\pi}}
 e^{-\al P_-}  P_-^{l+(3/2)} e^{ix^- P_-}  \Omlket \nn\\
&={ 2\sqrt{2} \pi^2 (2l+3)!! \over l! \sqrt{l+1} } z^{l+1} 
(1+ix^+)^{1/2}  \nn\\
&\quad \times  \left( x\xbar + x^+x^- + z^2 + 1 + i(x^+-x^-)\right) ^{-(l+(5/2))} \Omlket \period
\end{align}
The  probability distribution takes the form 
\begin{align}
| \Psi_l(z, x, \xbar,  x^-, x^+)|^2 
 &\propto {z^{2l+2} (1+(x^+)^2)^{1/2} \over \left[(x\xbar + x^+x^-+z^2+1)^2 + (x^+-x^-)^2 
\right] ^{l+(5/2) } } \period
\end{align}
The rough profile of this distribution is as follows.  First, it vanishes when any  one of the variables  becomes large in its magnitude and this occurs more rapidly for  higher $l$. 
 It also vanishes, like  $z^{2l+2}$,  near the boundary $z=0$.  On the other hand, 
it tends to a constant when $x$'s become small.  
The fall off of $\Psilket$ as $\sim z^{l+1}$ near the boundary can be 
understood more physically from  the Schr\"odinger  equation 
(\ref{Schrodeq}). The structure of the Hamiltonian $H^{l.c.}=i\mbP^-$ is 
rather similar to that of a system  in a centrifugal potential depending on the 
angular momentum $l$, such as the hydrogen atom.  Near  $z=0$, the requirement of the absence of singularity 
dictates  the wave function to be of the form $z^\al$, where 
$\al$ satisfies $-\al^2+2\al +l^2-1=0$. Thus for the normalizable solution we must have $\al = l+1$. 
\section{Discussions}
In this work, we have succeeded in quantizing a superparticle in 
the  $AdS_5 \times S^5$ background with RR flux exactly and obtained 
the complete spectrum of  one-particle states.  It is gratifying that the result 
 precisely agreed with that of  supergravity, although the method 
 is totally  different.  

There are two major directions for future research. One is the understanding of 
 the GKP-W relation from the first-quantized viewpoint. In this regard,  we note an important apparent difference  between the supergravity analysis and our analysis. 
The equations of motion for  the supergravity fields are second order in the 
derivative with respect to the $AdS$ coordinates and hence one 
 obtaines two solutions. One is normalizable  (under a  norm appropriate in field theory)  and corresponds to the propagating particle mode.  The other is non-normalizable and 
is thought to play the role of the source for the gauge-invariant 
super-Yang-Mills operator  placed  on the boundary of $AdS$. In contrast, 
 in our approach, the superconformal primary condition is an equation linear 
 in the derivative and hence we obtained  a unique solution as the highest weight 
state of the unitary representation. It is normalizable as a quantum mechanical 
 wave function.  It is not  clear to us at the moment whether  we should look for the  missing ``non-normalizable" states. One  reason is that the first quantized approach inherently  deals with a physical particle and hence non-particle mode may not be described. 
Another reason is that if we can construct the vertex operators anchored 
at points on the boundary which carry the quantum numbers of the corresponding 
particle modes,  it  should suffice to compute their correlation functions to see if they 
 corresond to those in the super-Yang-Mills theory.  If this is successful, one will not
 need the non-normalizable states, at least explicitly. In any case, construction of 
 the appropriate vertex operators will be a major goal   in this direction. 

Another important direction,  of course,  is the extension of our method to the 
superstring case. The first   task is the construction of the appropriately 
normal-ordered quantum superconformal generators. Once they are 
obtained,  one can start solving the superconformal primary conditions. 
Due to the presence of the non-zero modes,  the $su(4)$ part of the wave function 
 will not be unique in contrast to the particle case and this will  lead to 
 many solutions  for the superconformal primaries. 
Nevertheless we may hope that,  perhaps by devising some judicious ansatz,  at least
 some  of the solutions can be obtained.  It would be extremely interesting 
 if in such an attempt we  need to  discover some ``integrable structure" for the diagonalization of the  spectrum, just as in the super-Yang-Mills case. 

Some preliminary investigations in  these directions are
 underway and we hope to report our progress elsewhere. 
\par\bigskip\noindent
{\large\bf Acknowledgment}\par\smallskip\noindent
The research of T.H is supported in part by JSPS Research Fellowships for Young
Scientists, 
while the research of  Y.K. is supported in part by the
 Grant-in-Aid for Scientific Research (B)
No.~12440060  from the Japan
 Ministry of Education, Culture, Sports,  Science and Technology. 

\setcounter{equation}{0}
\renewcommand{\theequation}{A.\arabic{equation}}
\section*{{\bfall Appendix A:\ \  On the orbital $su(4)$ generator $l^i{}_j$ }}
In this appendix, we will elaborate on the orbital  $su(4)$ generator $l^i{}_j$, 
which plays  a crucial role in the analysis of the allowed representations of $psu(2,2|4)$ for our system. 

According to the general formula  (\ref{Noeformula}), 
the full $su(4)$ Noether charge  $\mbJ^i{}_j$ is given by 
\begin{align}
\mbJ^i{}_j &= {1\over e} (G^{-1} J^i{}_j G)^\una J_B^\una \period
\end{align}
Consider the part of $G^{-1} J^i{}_j G$ independent of the fermionic coordinates.  Since $g_x$ and $g_\phi$ commute with $J^i{}_j$,  it collapses to the 
purely  $su(4)$ expression 
 $g_y^{-1} J^i{}_j g_y$,  which takes value in $su(4)$. The orbital generator 
$l^i{}_j$ is then defined as
\begin{align}
l^i{}_j \equiv {1\over i e} (g_y^{-1} J^i{}_j g_y)^\Ap J_B^\Ap \period 
\end{align}
It is not difficult to check  that under the Dirac bracket $il^i{}_j$ 
 satsify the canonical $su(4)$ commutation relations 
$\Dcom{il^i{}_j}{il^k{}_n} = \delta^k_j (il^i{}_n) -\delta^i_n (il^k{}_j)$. 
Upon quantization, $l^i{}_j$ satisfies the same form of the algebra. 

Let us give a more explicit form of $l^i{}_j$. To this end, define 
a  $4\times 4$ matrix $U_t$ depending on a parameter $t$ as 
\begin{align}
U_t &\equiv \exp\left( t {i \over 2} y^\Ap \ga^\Ap \right) \period 
\end{align}
Then one can show the following relation 
\begin{align}
g_{-ty} J^i{}_j g_{ty} &= (U_{-t})^i{}_k J^k{}_l (U_t)^l{}_j \comma 
\label{gJg}
\end{align}
where $g_{ty} \equiv \exp( t y^i{}_j J^j{}_i )$. 
Since the equality is trivial at $t=0$, this can be proved by demonstrating 
 that  both sides satisfy the same first order differential equation 
 with respect to $t$.  Now set $t=1$ and substitute the expression 
 of $J^i{}_j$ in terms of the $SO(6)$ generators, namely 
\begin{align}
J^i{}_j &= {1\over 4}(\ga^{\Ap\Bp})^i{}_j J^{\Ap\Bp}+ {i \over 2} (\ga^\Ap)^i{}_j J^\Ap  \comma 
\end{align}
into the right hand side of (\ref{gJg}) and  focus on the coset part 
 proportional to $J^\Ap$. Then we obtain 
\begin{align}
(g^{-1}_y J^i{}_j g_y )^\Ap &= {i \over 2} (U^{-1} \ga^\Ap U)^i{}_j \comma 
\end{align}
where $U \equiv U_{t=1}$. 
Applying this formula  to  the definition of $l^i{}_j$  we get 
\begin{align}
l^i{}_j &= {1\over 2e}  (U^{-1} \ga^\Ap U)^i{}_j J^\Ap_B 
=  {1\over 2}  (U^{-1} \ga^\Ap U)^i{}_j 
{(\del J /\del \dot{X})^{-1}}_\Ap{}^\Bp P_\Bp \period \label{formlij}
\end{align}
In the second equality we expressed $J^\Ap_B$ in terms of 
 the phase space variables. This form of $l^i{}_j$ was utilized in the calculation of the Noether charges. Further, the explicit form of $U$ can be easily computed:
\begin{align}
U &= \cos \frac{|y|}{2} + i  {\gamma^\Ap y^{\Ap}
\over |y|}\sin \frac{|y|}{2}\comma 
\qquad |y| \equiv \sqrt{y^\Ap y^\Ap} \period  
\end{align} 

With the form of $l^i{}_j$ given in (\ref{formlij}), it is not difficult to prove the 
important quadratic identity (\ref{llid}).  First consider the classical case. Using 
(\ref{formlij}) we have
\begin{align}
l^i{}_j l^j{}_k &= {1\over 4e^2} (U^{-1} \ga^\Ap \ga^\Bp U)^i{}_k J^\Ap_B
J^\Bp_B  = {1\over 4e^2} \delta^i_k (J^\Ap_B J^\Ap_B) \period 
\label{llprod}
\end{align}
Taking the trace with respect to the indices $i,k$  we obtain
$\lhat^2  \equiv l^i{}_j l^j{}_i = {1\over e^2} (J^\Ap_B J^\Ap_B) $. Putting this back into (\ref{llprod}) we get the formula 
\begin{align}
l^i{}_jl^j{}_k = {1\over 4}\delta^i_k \lhat^2 \period  \label{llclass}
\end{align}
The quantum version of $l^i{}_j$ is given by the second part of 
 the formula (\ref{formlij}) with $P_\Bp$ understood as the differential  operator
$-i \del /\del y^\Bp$ and symmetrized as 
${1\over 4} (V^{i \Bp}_j  P_\Bp +  P_\Bp V^{i \Bp}_j )$, 
where $ V^{i \Bp}_j  =(U^{-1} \ga^\Ap U)^i{}_j {(\del J /\del \dot{X})^{-1}}_\Ap{}^\Bp$. This is needed  to realize the hermiticity property 
$(l^i{}_j)^\dagger = l^j{}_i$. 
Then due to the  re-ordering, quantum version of the formula (\ref{llclass}) acquires 
an extra term linear in $l^i{}_j$ and reads 
\begin{align}
l^i{}_jl^j{}_k = {1\over 4}\delta^i_k \lhat^2 +2l^i{}_j \period  
\end{align}
\setcounter{equation}{0}
\renewcommand{\theequation}{B.\arabic{equation}}
\section*{Appendix B:\ \  List of classical Noether charges }
In this appendix, we give the list of the classical Noether charges.  The following
notations are used: $(S_{\eta})^2 = S_{{\eta}^i}S_{{\eta}_i}, (S_{\theta})^2 = S_{{\theta}^i}S_{{\theta}_i}$. 
\begin{align}
\mathbb{P}^+ &  = iP_-, \\
\mathbb{P}^x & =  iP_{\bar{x}}, \\
\mathbb{P}^{\bar{x}} & =   iP_x, \\
\mathbb{P}^- & =  iP_+ = \frac{i}{4P_-}\left[ (-4P_xP_{\bar{x}} - e^{-2\phi}P_{\phi}^2
- e^{-2\phi}( l^2 + (S_{\eta}^2)^2 +
 4l^i{}_jS_{{\eta}^j}S_{{\eta}_i}) \right], \\
\mathbb{Q}^{+i} & =   -\sqrt{P_-}S_{\theta^i}, \\
\mathbb{Q}^+{}_i & =   \sqrt{P_-}S_{\theta_i}, \\
\mathbb{Q}^{-i} & =   \frac{1 }{2\sqrt{P_-}}\left[ - 2P_xS_{\theta^i}
+ iP_{\phi}e^{-\phi}S_{\eta^i} + 
e^{-\phi} \left( S_{\eta^i}(S_{\eta})^2
 + 2l^i{}_kS_{\eta^k} \right)  \right], \\
\mathbb{Q}^-{}_i & = - \frac{1}{2\sqrt{P_-}} \left[ - 2P_{\bar{x}}S_{\theta_i} 
- iP_{\phi}e^{-\phi}S_{\eta_i}
+ e^{-\phi} \left( S_{\eta_i}(S_{\eta})^2
  + 2S_{\eta_k}l^k{}_i  \right) \right], \\
\mathbb{S}^{+i} & =  \sqrt{P_-}\left( e^{\phi}S_{\eta^i} + i \bar{x}S_{\theta^i} \right)
 + i\tau\mathbb{Q}^{-i}, \\
 \mathbb{S}^+{}_i & = - \sqrt{P_-}\left( e^{\phi}S_{\eta_i} - ixS_{\theta_i} \right)
 - i\tau\mathbb{Q}^-{}_i, \\
 \mathbb{S}^{-i} & = \frac{1}{2\sqrt{P_-}}\left[ 2P_{\bar{x}}e^{\phi}S_{\eta^i} 
 + 2S_{\eta^i}S_{\theta^k}S_{\eta_k} 
 + S_{\theta^i} \left( (S_{\theta})^2  
   - iP_{\phi} \right) + 2l^i{}_kS_{\theta^k} \right]
 + i x^-\mathbb{Q}^{+i} + i  x\mathbb{Q}^{-i}, \\
 \mathbb{S}^-{}_i & = -\frac{1}{2\sqrt{P_-}}\left[ 2P_xe^{\phi}S_{\eta_i} 
 + 2S_{\eta_i}S_{\eta^k}S_{\theta_k} 
 + S_{\theta_i}\left( (S_{\theta})^2 
  + iP_{\phi} \right) + 2S_{\theta_k}l^k{}_i \right]
 - ix^-\mathbb{Q}^+{}_i - i\bar{x}\mathbb{Q}^-{}_i, \displaybreak[1]\\
 \mathbb{K}^+ & = \frac{1}{i} (e^{2\phi} + x\bar{x})P_- +  
 \tau(  iP_{\phi} + ixP_x + i\bar{x}P_{\bar{x}} + \tau \mathbb{P}^-), \\
 \mathbb{K}^x & = - ie^{2\phi}P_{\bar{x}}  
 +  x( iP_{\phi}  + ix^-P_- + ixP_x  + \frac{1 }{2}\left( -(S_{\theta})^2 
 + (S_{\eta})^2) \right)
-  ie^{\phi}S_{\theta^k}S_{\eta_k}
 - \tau\mathbb{J}^{-x} , \\
 \mathbb{K}^{\bar{x}} & = -i e^{2\phi}P_x
 + \bar{x}( iP_{\phi} +  ix^-P_- + i\bar{x}P_{\bar{x}}   + \frac{1 }{2}\left( (S_{\theta})^2
  - (S_{\eta})^2) \right)
  -i e^{\phi}S_{\eta^k}S_{\theta_k}
 - \tau\mathbb{J}^{-\bar{x}} , \\ 
  \mathbb{K}^- & = (x\bar{x} - e^{2\phi})\mathbb{P}^- 
  + x\mathbb{J}^{-\bar{x}} 
  + \bar{x}\mathbb{J}^{-x} 
  + ix^-P_{\phi} +  i(x^-)^2P_- 
  \nn\\
&\qquad 
  +\frac{i}{4P_-}\Bigl[ -(S_{\eta}{}^2)^2  + (S_{\theta}{}^2)^2
  + 4S_{\theta^k}S_{\eta_k}S_{\eta^l}S_{\theta_l} 
   \nn\\
&\qquad 
   + 4e^{\phi}(P_xS_{\theta^k}S_{\eta_k}
   + P_{\bar{x}}S_{\eta^k}S_{\theta_k}) 
   + 4l^k{}_l( S_{\theta^l}S_{\theta_k}
- S_{\eta^l} S_{\eta_k} ) \Bigr] , \displaybreak[1]\\
\mathbb{D} & = -iP_{\phi} - (i x^-P_- +  i xP_x +  i \bar{x}P_{\bar{x}}  )  - \tau \mathbb{P}^-
, \displaybreak[1]\\
\mathbb{J}^{+-} & = -ix^-P_- +  \tau\mathbb{P}^-, \\
\mathbb{J}^{x\bar{x}} & = -i\bar{x}P_{\bar{x}} + i xP_x + \frac{1 }{2} \left( (S_{\eta})^2 
- (S_{\theta})^2 \right), \\
\mathbb{J}^{+x} & = -ixP_- + i  \tau P_{\bar{x}}, \\
\mathbb{J}^{+\bar{x}} & = -i\bar{x}P_- + i \tau P_x, \\
\mathbb{J}^{-x} & = -x\mathbb{P}^- + i x^-P_{\bar{x}} 
+ \frac{ P_{\bar{x}}}{2P_-}\left( (S_{\eta})^2 
+ (S_{\theta})^2 \right) 
- \frac{1}{\sqrt{P_-}}S_{\theta^k}\mathbb{Q}^-_k
, \\
\mathbb{J}^{-\bar{x}} & = -\bar{x}\mathbb{P}^- + i x^-P_x 
- \frac{ P_x}{2P_-}\left( (S_{\eta})^2 + (S_{\theta})^2 \right)
- \frac{1}{\sqrt{P_-}}\mathbb{Q}^{-k}S_{\theta_k}
 , \\
 \mathbb{J}^i{}_j & =  l^i{}_j -  (S_{\eta^i}S_{\eta_j} 
 - \frac{1}{4}(S_{\eta})^2\delta^i{}_j) -  (S_{\theta^i}S_{\theta_j} - 
 \frac{1}{4}(S_{\theta})^2\delta^i{}_j).
 \end{align}
\setcounter{equation}{0}
\renewcommand{\theequation}{C.\arabic{equation}}
\section*{Appendix C:\ \  Analysis of the allowed {\boldmath $su(4)$} representations }
In this appendix, we provide some details of the analysis of the allowed 
$su(4)$ representations sketched in the main text. 

Let us first describe the analysis of the relations (\ref{JJrel}) valid on superconformal 
primaries. Just as in the analysis of the relation (\ref{llrel}) for the orbital part, 
 $\calJ^2{}_1 \approx 0$, $\calJ^3{}_2 \approx 0$ and $\calJ^4{}_3 \approx 0$ give the following 3 equations:
\begin{align}
(i) \quad & (\lam_1 + 2\lam_2 + \lam_3  + N-2 ) E^-_1 \ket{\lam_1, \lam_2, \lam_3}  =0 \comma \\
(ii) \quad & (\lam_1-\lam_3 + 4-N ) E^-_2 \ket{\lam_1, \lam_2, \lam_3}  =0 \comma \\
(iii) \quad & (\lam_1 + 2\lam_2 + \lam_3 + 6-N) E^-_3 \ket{\lam_1, \lam_2, \lam_3}   =0 \period
\end{align}
Here and hereafter, $\ket{\lam_1, \lam_2,  \lam_3}$ refers to the 
$su(4)$ highest weight state  in the total Hilbert space $\calH_{tot} =
\calH_{spin} \otimes 
\calH_{orb}$, consisting   of the spin part and the orbital part. 
When we need to emphasize this feature, we will denote the state 
as $\ket{\lam_1, \lam_2,  \lam_3}_{tot}$. 

Consider the equation $(i)$. 
For $N \ge 3$ the coefficient is non-vanishing and we must have 
$E^-_1 \ket{\lam_1, \lam_2, \lam_3}  =0$ and hence $\lam_1=0$. 
Similarly, from $(iii)$ we find  $\lam_3=0$ for $N \le 5$.  Therefore   for $3 \le  N \le  5$ the relation  $(ii)$ reduces to 
$(4-N)  E^-_2 \ket{0, \lam_2, 0}  =0$. This tells us that   $\lam_2$ is arbitrary for $N=4$, while for $N=3,5$ only  the singlet state $\ket{0,0,0}$ is allowed. 

Next consider the cases with $N \le 2$. We already know that $\lam_3=0$. Thus $(ii)$ becomes $(\lam_1 + 4-N)E^-_2 \ket{\lam_1, \lam_2, 0} =0$. But since $4-N >0$, we must have $E^-_2 \ket{\lam_1, \lam_2, 0} =0$ and hence $\lam_2=0$. 
Then $(i)$ reduces to $ (\lam_1 + N-2) E^-_1 \ket{\lam_1, 0,0} =0$. 
From this  we easily find  that the possible values of $\lam_1$ are  $\lam_1=0$ for $N=2$, $\lam_1=0,1$ for $N=1$, 
and $\lam_1=0,2$ for $N=0$. Actually  for $N=0$  only the singlet $\ket{0,0,0}$  is allowed  since without exciting any fermionic oscillator we cannot produce the state  $\ket{2,0,0}$.  Since the analysis for the cases  with $N \ge 6$ is very similar, 
 it will be omitted. 

Combining these  results,  one finds that 
at this point the following highest weight states are allowed:
 $\ket{0,0,0}_{tot}$ for $N=0,1, 2,3, 5, 6, 7, 8$, $\ket{1,0,0}_{tot}$ for $N=1$, 
$\ket{0,0,1}_{tot}$ for $N=7$  and $\ket{0,\lam, 0}_{tot}$ with 
 arbitrary non-negative integer  $\lam$ for $N=4$.  

These states must be realized 
 as the tensor products  of the spin part and the orbital part.  The spin part is generated
by the fermionic oscillators $S^i$ and $\Stil^i$. From the form of the 
spin part of the $su(4)$ generators given in (\ref{spingen}),  one can easily find 
the Dynkin labels carried by $S^i$ and $\Stil^i$. They 
are  $[1,0,0]$ for $S^1, \Stil^1$, $[-1,1,0]$ for 
$S^2, \Stil^2$, $[0,-1,1]$ for $S^3, \Stil^3$ and $[0,0,-1]$ for 
$S^4, \Stil^4$. As for the orbital part, we already know that the allowed 
highest weight states are of the form $\ket{0,l,0}_{orb}$, with an arbitrary 
 non-negative integer $l$. With this information, one can easily analyze the 
possible value of $l$ for realizing the states  $\ket{0,0,0}_{tot}$, 
$\ket{1,0,0}_{tot}$, $\ket{0,0,1}_{tot}$  and $\ket{0,\lam, 0}_{tot}$
for each  relevant value of $N$. 

 For example, for $N=0$, {\it i.e} 
without exciting any fermionic oscillators,  the orbital part must be $\ket{0,0,0}_{orb}$
in order to realize $\ket{0,0,0}_{tot}$. As another example, consider the 
realization of the state $\ket{0,0,1}_{tot}$ for $N=1$. With one fermion excited 
 the highest weight state of the spin part carries   the Dynkin index $[1,0,0]$. 
Thus to produce $\ket{1,0,0}_{tot}$, the orbital part must again 
 be $\ket{0,0,0}_{orb}$. A slightly  non-trivial example occurs for $N=2$. 
With two fermionic oscillators excited, the spin part can be $S^1S^2 \ket{0}$, 
which is the highest weight state with the index $[0,1,0]$. Thus it can 
produce $\ket{0,0,0}_{tot}$ when tensored with $\ket{0,1,0}_{orb}$. So in this 
case  $l=1$  is allowed.  

As a result of this type of analysis, one obtains a more refined information 
 for the allowed highest weight states.  Let us summarize the result by 
listing the value of $N$, the allowed total highest weight, and the value of 
$l$ for  its orbital part:
\begin{align}
&N=0:\ \ket{0,0,0}_{tot},\ \  l=0\comma   \qquad \quad 
N=8:\ \ket{0,0,0}_{tot}, \ \ l=0 \comma \nn\\
& N=1:\ \ket{1,0,0}_{tot},\ \ l=0\comma  \qquad \quad 
 N=7:\ \ket{0,0,1}_{tot}, \ \ l=0 \comma \nn\\
& N=2:\ \ket{0,0,0}_{tot}, \ \ l=1 \comma \qquad \quad 
N=6:\ \ket{0,0,0}_{tot}, \ \ l=1 \comma \label{statelist} \\
 &N=3:\ \mbox{no solution} \comma \qquad \qquad \quad \ 
 N=5:\ \mbox{no solution} \comma \nn\\
& N=4:\ \ket{0,\lam_2, 0}_{tot}, \ \ l-2 \le \lam_2 \le l+2\comma 
\ \ \mbox{any  $l$} \period \nn
\end{align}

To further reduce the possibilities, one can impose the condition of unitarity for the 
representation of $psu(2,2|4)$ built upon these $su(4)$ states.  As mentioned in the 
main text, one can obtain several different bounds depending on the choice of the 
 pair of supercharges.  A particularly useful combination is the 
bound 
\begin{align}
E \ge \lam_1 + \lam_2 +\lam_3 \comma \label{utbound}
\end{align}
 which is powerful enough to eliminate many of the possible states. 

The information given in the list (\ref{statelist}) is sufficient to compute 
the energy of the superconformal primaries based on these $su(4)$ highest weight states. From (\ref{Jhatsq}) the energy can 
 be expressed as 
\begin{align}
\mbE &= {1\over 4} ( \mbJhat^2 -\lhat^2) + {1\over 16} (N-4)^2 -1 \period
\end{align}
The general formula for the value of the Casimir operator  $\mbJhat^2$ on the highest weight state $\ket{\lam_1, \lam_2, \lam_3}_{tot}$ reads\footnote{This formula can be easily derived 
 by using the exression of $\mbJ^i{}_j$ in the Chevalley basis, just as 
 in (\ref{Chevalley}). }
\begin{align}
&\mbJhat^2 \ket{\lam_1, \lam_2, \lam_3}_{tot} \nn\\
&\quad = \left( {1\over 4} (3\lam_1^2 + 2 \lam_1 \lam_3+3\lam_3^2) + (\lam_2 +3)(\lam_1 + \lam_3) + \lam_2 (\lam_2+4)  \right) 
\ket{\lam_1, \lam_2, \lam_3}_{tot} \period \label{ValJhatsq}
\end{align}
The  value of $\lhat^2$ on $\ket{0,\lam ,0}_{orb}$  was already quoted 
 in the main text to be $\lam(\lam+4)$, which is actually a special case of (\ref{ValJhatsq}).  Using these formulas, we can readily compute 
the AdS energy of the superconformal primary state 
which can be built upon the states listed above.  If we denote  the energy 
for the state with $N$ fermionic oscillators excited by $E_N$,  the result is 
\begin{align}
E_0 &=E_8 = 0 \comma \quad E_1 =E_7 = \half\comma \quad E_2 = E_6=-2 \comma \\
E_4 &= {1\over 4} \left( \lam_2(\lam_2+4) -l(l+4) \right) -1 \period
\end{align}
On the other hand, the bounds following from (\ref{utbound}) are,   $E\ge 0$ for $N=0,2,6, 8$, $E \ge 1$ for $N=1,7$,  and $E \ge \lam_2$ for $N=4$. Evidently, 
the cases for  $N=1,2,6,7$  are excluded, while the cases for  $N=0,8$ are allowed. 
As for the case with $N=4$, it is easy  to see that the bound 
 $E_4 \ge \lam_2$
 reduces  to $\lam_2 \ge l+2$ and  hence it is allowed for  $\lam_2=l+2$. 
 One can check that these allowed cases actually meet all the other bounds as well. 
\newpage

\end{document}